\documentclass[twocolumn]{aastex631}

\shorttitle{KIC Stellar Parameters}
\shortauthors{Zhang et al.}

\graphicspath{{./}{figures/}}

\begin{document}

\title{Photometric Stellar Parameters for 195,478 Kepler Input Catalog (KIC) Stars}

\correspondingauthor{Yang Huang}
\email{huangyang@ucas.ac.cn}

\author[0009-0008-7479-0742]{Bowen Zhang}
\affiliation{Key Lab of Optical Astronomy, National Astronomical Observatories, Chinese Academy of Sciences, Beijing 100101, People's Republic of China}
\affiliation{School of Astronomy and Space Science, University of Chinese Academy of Sciences, Beijing 100049, People's Republic of China}

\author[0000-0003-3250-2876]{Yang Huang}
\affiliation{School of Astronomy and Space Science, University of Chinese Academy of Sciences, Beijing 100049, People's Republic of China}
\affiliation{Key Lab of Optical Astronomy, National Astronomical Observatories, Chinese Academy of Sciences, Beijing 100101, People's Republic of China}

\author[0000-0003-4573-6233]{Timothy C. Beers}
\affiliation{Department of Physics and Astronomy and JINA Center for the Evolution of the Elements (JINA-CEE), University of Notre Dame, Notre Dame, IN 46556, USA}

\author[0000-0001-8424-1079]{Kai Xiao}
\affiliation{School of Astronomy and Space Science, University of Chinese Academy of Sciences, Beijing 100049, People's Republic of China}

\author[0000-0002-2874-2706]{Jifeng Liu}
\affiliation{Key Lab of Optical Astronomy, National Astronomical Observatories, Chinese Academy of Sciences, Beijing 100101, People's Republic of China}
\affiliation{School of Astronomy and Space Science, University of Chinese Academy of Sciences, Beijing 100049, People's Republic of China}

\author{Lei Jia}
\affiliation{Key Lab of Optical Astronomy, National Astronomical Observatories, Chinese Academy of Sciences, Beijing 100101, People's Republic of China}

\author[0000-0003-3474-5118]{Henggeng Han}
\affiliation{Key Lab of Optical Astronomy, National Astronomical Observatories, Chinese Academy of Sciences, Beijing 100101, People's Republic of China}

\author[0000-0002-7598-9250]{Zhirui Li}
\affiliation{Key Lab of Optical Astronomy, National Astronomical Observatories, Chinese Academy of Sciences, Beijing 100101, People's Republic of China}
\affiliation{School of Astronomy and Space Science, University of Chinese Academy of Sciences, Beijing 100049, People's Republic of China}

\author[0009-0006-7556-8401]{Chuanjie Zheng}
\affiliation{Key Lab of Optical Astronomy, National Astronomical Observatories, Chinese Academy of Sciences, Beijing 100101, People's Republic of China}
\affiliation{School of Astronomy and Space Science, University of Chinese Academy of Sciences, Beijing 100049, People's Republic of China}

\author[0000-0002-3935-2666]{Yongkang Sun}
\affiliation{Key Lab of Optical Astronomy, National Astronomical Observatories, Chinese Academy of Sciences, Beijing 100101, People's Republic of China}
\affiliation{School of Astronomy and Space Science, University of Chinese Academy of Sciences, Beijing 100049, People's Republic of China}

\author[0009-0000-6932-7894]{Ruifeng Shi}
\affiliation{School of Astronomy and Space Science, University of Chinese Academy of Sciences, Beijing 100049, People's Republic of China}

\author[0009-0007-5610-6495]{Hongrui Gu}
\affiliation{Key Lab of Optical Astronomy, National Astronomical Observatories, Chinese Academy of Sciences, Beijing 100101, People's Republic of China}
\affiliation{School of Astronomy and Space Science, University of Chinese Academy of Sciences, Beijing 100049, People's Republic of China}

\begin{abstract}

The stellar atmospheric parameters and physical properties of stars in the Kepler Input Catalog (KIC) are of great significance for the study of exoplanets, stellar activity, and asteroseismology. However, despite extensive effort over the past decades, accurate spectroscopic estimates of these parameters are available for only about half of the stars in the full KIC catalog. In our work, by training relationships between photometric colors and spectroscopic stellar parameters from Gaia DR3, the Kepler Issac-Newton Survey, LAMOST DR10, and APOGEE DR17, we have obtained atmospheric-parameter estimates for over 195,000 stars, accounting for 97$\%$ of the total sample of KIC stars. We obtain 1$\sigma$ uncertainties of 0.1\,dex on metallicity [Fe/H], 100\,K on effective temperature $T_{\rm eff}$, and 0.2\,dex on surface gravity log $g$. In addition, based on these atmospheric parameters, we estimated the ages, masses, radii, and surface gravities of these stars using the commonly adopted isochrone-fitting approach. External comparisons indicate that the resulting precision for turn-off stars is 20$\%$ in age; for dwarf stars, it is 0.07 $M_{\odot}$ in mass, 0.05 $R_{\odot}$ in radius, and 0.12 dex in surface gravity; and for giant stars, it is 0.14 $M_{\odot}$ in mass, 0.73 $R_{\odot}$ in radius, and 0.11 dex in surface gravity.

\end{abstract}
\keywords{Metallicity (1031), Stellar age (1581), Fundamental parameters of stars (555)}

\section{Introduction} \label{sec:intro}

As the most powerful and successful exoplanet explorer to date, the Kepler/K2 mission \citep{2010Sci...327..977B,2014PASP..126..398H} has discovered more than 3,300 exoplanets, over half of the currently confirmed exoplanets. The extremely precise and short-cadence ($\sim 1$ minute) light curves obtained from Kepler/K2 have revolutionized the field of asteroseismology. This data has facilitated the detection of Solar-like oscillations in over 500 main-sequence and subgiant stars \citep{2014ApJS..210....1C} as well as for more than twenty thousand red giants (e.g.,  \citealt{2013ApJ...765L..41S,2016MNRAS.463.1297Y}). 
This extensive dataset has enabled accurate modeling of the fundamental properties of these stars across nearly the entire low-mass HR diagram \citep{2020svos.conf..457H}. There are also 75 exoplanets confirmed from Kepler's archival data, demonstrating the enduring impact and vitality of this space mission. 

During its first four years of operations, Kepler made long-term observations of the Kepler field, a 116 square degree area of sky located between the constellations Cygnus and Lyra. To guide these observations, a catalog of some 200,000 stars, known as the \emph{Kepler Input Catalog} (KIC; \citealt{2011AJ....142..112B}), was determined in advance. 
Observations of this area accumulated a large amount of photometric data over temporal baselines on the order of years, which have not only greatly advanced the field of exoplanet search and characterization, but also provided an important basis for research in many other fields, including stellar activity, astroseismology, and the study of star clusters \citep{2010PASP..122..131G,2013ApJS..209....5S,2014ApJS..211...24M,2016RPPh...79c6901B}.

The fundamental stellar parameters of stars play a important role in refining our understanding of stellar theoretical models and evolution. In order to construct the appropriate stellar models to constrain their evolution, the physical properties of stars, in particular the metallicity, are essential input parameters \citep{2004ApJ...613..898T,2012MNRAS.427..127B}. The situation is similar for the estimation of stellar ages and masses; a precise metallicity estimate of a star is required. After determining the stellar age, the age-rotation relation can be analyzed to a high level of precision 
\citep{2020A&A...634L...9W,2022ApJ...937...94M}. The metallicity also deeply influences the stellar atmosphere and structure, as well as the relationship between stellar activity and metallicity \citep{2018ApJ...852...46K,2023MNRAS.524.5781S,2024ApJ...970...53L}.

The nature of the exoplanet(s) associated with a star is expected to be related to the physical properties of the host. The probability that a star hosts a planet, as well as the type of the planet, are both influenced by the elemental abundances of the proto-planetary disk \citep{1997MNRAS.285..403G,2010PASP..122..905J,2018AJ....155...89P,2024arXiv240713821B}. Furthermore, the radius gap, a region which shows a deficit of planet occupation in the planet radius-mass map (at around 1.9 $R_{\oplus}$), is also thought to be influenced by the host star's metallicity, mass, and age \citep{2013ApJ...776....2L,2017ApJ...847...29O,2020ApJS..247...28H,2024NatAs...8..463B,Yun2024}.

However, we still lack full information on the fundamental stellar parameters of KIC stars, particularly metallicity, which is crucial for accurately determining other parameters as mentioned above.
Attempts to obtain this information continued throughout the Kepler mission, both before and after. 
Prior to the launch of the Kepler satellite, the atmospheric parameters of KIC stars were estimated through the use of broad-band photometry. For example, \citet{2011AJ....142..112B} provided estimates of the metallicity, effective temperature, surface gravity, and extinction toward KIC stars using a Bayesian posterior estimation method based on this photometry. However, the stellar parameters predicted by this method differ significantly from those obtained from both low-resolution (e.g., \citealt{2014ApJ...789L...3D}) and high-resolution spectroscopic studies with the KECK telescope (e.g., \citealt{2017AJ....154..108J}).

The \emph{Kepler Stellar Properties Catalog} (KSPC; \citealt{2014ApJS..211....2H}) provided revised stellar parameters for 138,600 targets in Quarters 1–16 (Q1-16), using colors, proper motions, spectroscopy, parallaxes, and Galactic population-synthesis models. However, only 7\% stars in this catalog had spectroscopic information at that time. Applying similar methods, \citet{2017ApJS..229...30M} provided  stellar parameters for 197,096 targets in Quarters 1–17 (Q1-17). Again, spectroscopic parameters were available for no more than 10\% sample stars. Most recently, this group \citep{2020AJ....159..280B} provided a catalog for 186,301 Kepler Stars, with fundamental properties (including stellar ages) homogeneously estimated from isochrone fitting using broadband photometry, and Gaia Data Release 2 parallaxes, as well as spectroscopic metallicities if available. Compared to previous versions, the fraction of stars with spectroscopic information has increased to approximately 35\% (about 66,000 stars), less than half of the number of KIC stars.

Another major effort for obtaining stellar parameters for KIC stars are large-scale spectroscopic surveys, including the LAMOST-Kepler Survey 
\citep{2015ApJS..220...19D,2018ApJS..238...30Z,2020RAA....20..167F,2022SSPMA..52B9502F}, the APOGEE-Kepler Survey \citep{2014ApJS..215...19P,2018AAS...23145013P}, and the California-Kepler Survey \citep[CKS;][]{2017AJ....154..107P, 2017AJ....154..108J, 2018AJ....155...89P}.
As shown in Table \ref{tab:crossmatch}, these three surveys have obtained stellar parameters for 78,141, 23,198, and 1716 stars by cross-matching their data releases with the KIC catalog.
Despite these extensive efforts, the total number of stars with spectroscopically measured atmospheric parameters is 85,986, still less than 50\% of the number of KIC stars.

More recently, stellar atmospheric and other physical parameters have been derived using narrow- or medium-band photometric surveys, particularly those in the near-ultraviolet bands \citep[e.g.,][]{2015ApJ...803...13Y, 2019ApJS..243....7H, 2021ApJS..254...31C, 2022ApJ...925..164H, 2023ApJ...957...65H, 2024arXiv240802171H}.
The precision is comparable to that achieved from low- or medium-resolution spectroscopy.
In recent decades, the Kepler field has been observed using near-ultraviolet bands, such as the Kepler-INT Survey \citep[KIS;][]{2012AJ....144...24G,2012arXiv1212.3613G}, or through the use of Gaia XP spectra. Parameter-sensitive narrow- or medium-band photometric colors (for example the well-known Str{\"o}mgren filter system) can be readily integrated from the flux-calibrated low-resolution spectra of the latter survey.  In this work, we aim to determine stellar parameters for all KIC stars using narrow- or medium-band photometric colors from these surveys, with spectroscopic labels from the LAMOST serving as training data.
The paper is structured as follows. Section \ref{sec:Data} describes the data. Section \ref{subsec:extinction} presents the 3-D extinction map toward the Kepler field. Atmospheric parameters are determined in Section \ref{subsec:Parameters}, while physical parameters are estimated in Section \ref{subsec:Isofit}. Finally, Sections \ref{subsec:Discussion} and \ref{subsec:conclusion} provide a discussion and a summary.

\begin{deluxetable}{cc}[h!]
\tablenum{1}
\tablecaption{KIC Stars in Large-scale/dedicated Spectroscopic Surveys\label{tab:crossmatch}}
\tablewidth{0pt}
\tablehead{
\colhead{Catalog/Survey} & \colhead{Number}
}
\decimals
\startdata
KIC & 200,038\\
LAMOST DR10 & 79,015\\
APOGEE DR17 & 23,198\\
CKS DR1 $\&$ DR2& 1716\\
KIC stars with spectroscopic parameters & 86,482\\
\enddata
\end{deluxetable}

\vskip 2cm
\section{Data} \label{sec:Data}

The Kepler field, a 116 square-degree region situated between Cygnus and Lyra, is centered at celestial coordinates $(\alpha,~\delta)=(\text{290}^{\circ}$,~$\text{45}^{\circ})$ and Galactic coordinates  $(l,~b)=(\text{76}^{\circ}$,~$\text{14}^{\circ})$ \citep{2011AJ....142..112B}, and has attracted a multitude of surveys. In this work, we primarily employed data from Gaia Data Release 3, the Kepler-INT Survey Data Release 2, the Large Sky Area Multi-Object Fiber Spectroscopic Telescope Data Release 10, and the Apache Point Observatory Galactic Evolution Experiment Data Release 17, supplemented with distance estimates sourced from the catalog by \cite{2021AJ....161..147B}.

\subsection{Gaia DR3}

The third Gaia data release (Gaia DR3) \citep{2023A&A...674A...1G}, derived from observations over a 35-month period, not only includes low-resolution ($R = \lambda / \Delta \lambda \sim 50$) BP/RP (XP) spectra for around 220 million sources, predominantly those with magnitudes brighter than $G<17.65$ (well-calibrated both internally and externally by \citealt{2021A&A...652A..86C}, \citealt{2023A&A...674A...2D}, and \citealt{2023A&A...674A...3M}, respectively), but also offers the most precise photometric data $(G,~BP,~RP)$ to date for approximately 1.8 billion stars \citep{2021A&A...649A...1G,2021A&A...650C...3G,2021A&A...649A...3R}. This provides the high-quality photometric and slitless spectroscopic data essential for conducting our study.

Based on the Gaia XP spectra, we further synthesized Str\"omgren photometry for the $vby$-bands using the generation function provided by the Python 
package $\texttt{GaiaXPy}$ \citep{2022gaiaxpy}.

 \subsection{KIS DR2}
The Kepler field is observed by the Kepler-INT Survey \citep[KIS;][]{2012AJ....144...24G,2012arXiv1212.3613G}, which employs the Isaac Newton Telescope (INT) to collect photometric data. The second KIS data release (DR2) includes $U$-, $g$-, $r$-, $i$-, and $H_{\rm \alpha}$-band photometry for 14.5 million stars, spanning a 113 square degree area of the Kepler field. 
In particular, the near-violet $U$-band photometry in KIS DR2 is crucial for the analysis presented in this work.

 \subsection{LAMOST DR10}

The Large Sky Area Multi-Object Fiber Spectroscopic Telescope (LAMOST) features a unique quasi-meridian reflecting Schmidt design outfitted with 4000 optical fibers, covering a 20 square degree field-of-view. Its tenth Data Release (DR10) provides an extensive collection of tens of millions of low-resolution ($R \sim 1800$) spectra across the optical spectrum from 3800 to 9000\,\AA. For estimation of the primary stellar atmospheric parameters (effective temperature $T_{\rm eff}$, surface gravity ${\rm log}$~$g$, and metallicity $\rm [Fe/H]$), the project relies on the LAMOST Stellar Parameter Pipeline for AFGK-stars \citep[LASP;][]{2011RAA....11..924W,2014IAUS..306..340W} and LAMOST Stellar Parameter Pipeline for M-stars \citep[LASPM;][]{2021RAA....21..202D}. 

 \subsection{APOGEE DR17}

The Galactic Evolution Experiment at Apache Point Observatory (APOGEE), as a part of the SDSS-III initiative, aimed to comprehensively address galaxy formation by conducting an unprecedented large-scale survey with detailed chemical and kinematic analysis. The APOGEE Stellar Parameter and Chemical Abundances Pipeline \citep[ASPCAP;][]{2016AJ....151..144G} delivers high-precision estimates of stellar parameters, including effective $T_{\rm eff}$, surface gravity ${\rm log}$~$g$, and metallicity $\rm [Fe/H]$. APOGEE DR17 published stellar atmosphere parameters for about 0.73 million stars, achieving measurement precision of typically 2\%, 0.1\,dex, and 0.05\,dex for $T_{\rm eff}$, log $g$, and $\rm [Fe/H]$, respectively.

In this work, we utilized both photometric and spectroscopic data to estimate the atmospheric parameters of KIC stars. To achieve this, we cross-matched the KIC catalog with data from the aforementioned surveys. After cross-matching, we found 199,571 stars with Gaia DR3 ultra wide-band photometry, 197,157 stars with Gaia DR3 XP spectra, and 190,604 stars with KIS DR2 photometry. For spectroscopic data, we found 23,198 stars observed by APOGEE and 79,015 by LAMOST, as summarized in Table\ref{tab:crossmatch}. To incorporate extinction values from the map derived in Section \ref{subsec:extinction}, we required distance estimates for the KIC stars. Cross-matching with \citet{2021AJ....161..147B} provided distance information for 197,064 stars. All cross-matching processes were conducted using TopCat \citep{2005ASPC..347...29T}, employing the best-matching model and a 3-arcsecond matching radius.

\vskip 1cm
\section{Construction of a 3-D Dust Map for the Kepler field with the 'Star-Pair' Method}\label{subsec:extinction}

In this study, the \cite{1998ApJ...500..525S} dust map $E(B-V)$ is not utilized for reddening correction, due to its inadequacies at low Galactic latitudes and the presence of spatially dependent errors, as reported in recent work by Sun et al. (submitted). Instead, the 3-D dust map for the Kepler field, derived through the straightforward `star-pair' (hereafter SP) method \citep[][see their Section 5 for more details]{2013MNRAS.430.2188Y} is employed. 

The central idea behind the SP method is that stars with similar atmospheric parameters -- metallicity, effective temperature, and surface gravity -- exhibit analogous intrinsic colors. The SP method typically involves defining the relationship between the intrinsic colors and the physical quantities using a sample of low-extinction stars, which is then applied to the entire sample to obtain $E(BP-RP)$. 

A detailed description of the SP method with the Gaia DR3 photometry color $BP-RP$ and LAMOST DR10 spectroscopic stellar parameters is as follows.

\begin{itemize}

\item We combine the Gaia DR3 photometric data with the spectroscopic data from LAMOST DR10, as well as the \cite{2021AJ....161..147B} distance catalog, with cross-matching radius of $3''$. A reference sample, constituting 1,037,145 stars, is selected with the following constraints: 1) Signal-to-noise ratio for the $g$-band (${\rm SNR}_{\rm g}$) of the LAMOST spectra greater than $20$; 2) Galactic latitude higher than $40^{\circ}$; 3) The 3-D dust map from \cite{2019ApJ...887...93G}, represented as $E(B-V)_{\rm G19}$, is less than 0.01; 4) \cite{2021AJ....161..147B} relative distance error less than 30\%, in order to avoid poorly constrained distance information.

\item To construct the Kepler field target sample, we combine the Gaia DR3 photometric data with the spectroscopic data from LAMOST DR10, as well as the \cite{2021AJ....161..147B} distance catalog, with a cross-matching radius of $3''$. The target sample includes 126,277 stars that meet the following constraints: 1) SNR$_{\rm g}$ of the LAMOST spectra more than $20$; 2) Located in the sky area whose RA ranging from 279$^\circ$ to 302$^\circ$ and DEC ranging from 36$^\circ$ to 52$^\circ$; 3) \cite{2021AJ....161..147B} relative distance error less than 30\%.

\item The $BP-RP$ is adopted from the Gaia $BP$ and $RP$ bands; the intrinsic color
$(BP-RP)_0$ can be estimated from $(BP-RP)_0=BP-RP-E(BP-RP)$. To obtain the reddening value, $E(BP-RP)$, a transformation is performed as shown in the following equation:
\begin{equation}
E(BP-RP)=(R_{\rm BP}-R_{\rm RP}) E(B-V)_{\rm G19}~,
\label{equ:EBPRPEBV}
\end{equation}
where $R_{\rm BP/RP}$ is the reddening coefficient with respect to $E(B-V)_{\rm G19}$ for the $BP$- and $RP$-bands, respectively, which can be calculated with:
 \begin{equation}
R_{\rm BP/RP} =R_{\rm V} (A_{\rm BP/RP}/A_{\rm V})~,
\label{equ:Rxp}
\end{equation}
\noindent where $R_{\rm V}$ represents the total-to-selective extinction ratio, defined as $R_V \equiv \frac{A_V}{E (B - V)}$. Here, instead of fixing it at 3.1, we use the actual measurements from \cite{2023ApJS..269....6Z} for each target. The $A_{\rm BP/RP}$ and $A_{\rm V}$ are the reddening values in the $BP/RP$- and $V$- bands, respectively. The reddening ratio of $A_{\rm BP/RP}/A_{\rm V}$ is taken from \cite{2019ApJ...877..116W}.

For each target star, the reference stars are selected from the reference sample as those having values of $T_{\rm eff}$, ${\rm log}~g$, and $\rm [Fe/H]$ that differ from those of the target by less than 130\,K, 0.06\,dex, and 0.06\,dex, respectively. The box sizes for selecting reference stars are empirically determined to ensure both a sufficient number of stars and a clear relationship between intrinsic color and atmospheric parameters within the box range. The extinction values for the target stars $E(BP-RP)$ are measured from the difference between the observed color $BP-RP$ and intrinsic color $(BP-RP)_0$. The latter is derived both assuming that the intrinsic colors of the target and its control stars vary linearly with $T_{\rm eff}$, ${\rm log}~g$, and $\rm [Fe/H]$, and based on the random forest machine-learning fitting technique \citep{2001MachL..45....5B}. From comparison with the results of the above two techniques, the outcome of the random forest approach has been selected as the final result.

\item To construct a continuous 3-D extinction map applicable to all KIC stars, it is required to interpolate the discrete reddening values we have obtained. We subdivided the Kepler field into a grid of $10^{\prime}\times 10^{\prime}$ squares, assuming that the stars within each grid share the same line of sight. For the stars in each grid square, we employed two interpolation techniques: cubic-function fitting and Gaussian- error function fitting. We then derived continuous color-excess values. Among these two interpolation strategies, the method demonstrating superior goodness of fit, as measured by the coefficient of determination, was adopted as the final choice.
\end{itemize}

\begin{figure*}[ht!]
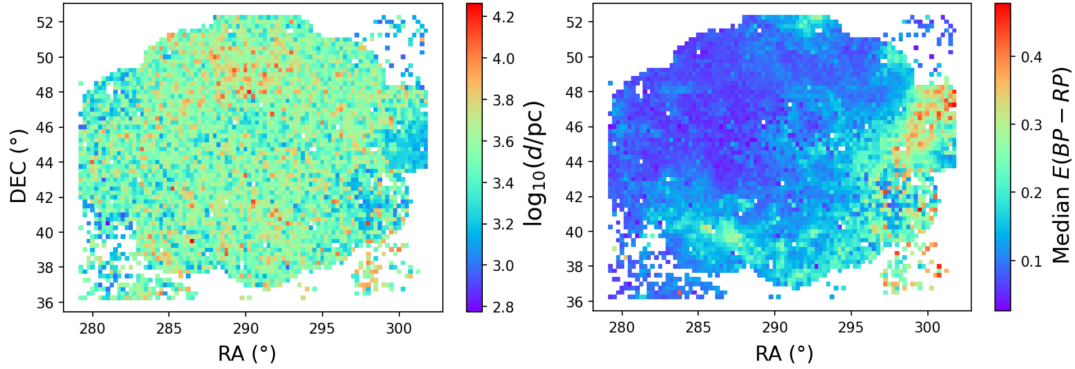

\gridline{\fig{/starpair/depth.png}{0.8\textwidth}{}
}

\caption{Depth (left panel) and median $E(B-V)$ (right panel) of the 3-D reddening map toward Kepler field, coded by the color bars shown to the right of each panel.
\label{fig:depth&median}}
\end{figure*}

Finally, we constructed a 3-D extinction map with an angular resolution of 10 arcmin and a distance resolution of 20 parsecs in the Kepler field, as shown in Figure \ref{fig:depth&median}.
As a first check, the extinction values $E(BP-RP)$ yielded by the SP method is directly compared to those from $E(B-V)_{\rm G19}$ (see Figure \ref{fig:compareEBV}). Generally, they are very consistent with each other, with a negligible offset and a moderate scatter of 0.037 mag. We further assess the accuracy of reddening derived from the SP technique using member stars of open clusters, where extinction values are assumed to be constant. In the Kepler field, there are four open clusters: NGC 6811, NGC 6819, NGC 6866, and NGC 6791. Using positions, distances, and proper motions from Gaia DR3 \citep[following the methods in][]{2019ApJS..243....7H, 2023ApJ...944...88L}, we selected member stars of these four open clusters according to their mean positions, distances, and proper motions reported in \cite{2021AJ....161..147B, 2023AA...673A.114H}. The extinction distributions $E (B - V)$ for the member stars, derived both from our SP technique and from \cite{2019ApJ...887...93G}, are shown in Figure~\ref{fig:interpolation}.
It is evident that the SP technique produces narrower distributions for all four clusters, indicating that the internal precision of the SP method is significantly higher than that of \cite{2019ApJ...887...93G}.
Typically, the scatter in the extinction distributions obtained using the SP technique is significantly smaller than 0.01 mag, whereas the scatter from \cite{2019ApJ...887...93G} exceeds 0.025 mag. The median extinction values for the member stars of all four clusters are consistent with those reported in the literature \citep[see Table \ref{tab:openCluster},][]{2013AJ....145....7J,2014AJ....148...51A,2019AA...623A.108B}.

\section{Stellar Atmospheric-parameter Estimation}\label{subsec:Parameters}\

Multi-band photometric data, particularly if narrow-band filters are involved, provide an extremely efficient means to estimate stellar-atmospheric parameters, as has been recognized for over half a century.  For example, \citet{1963QJRAS...4....8S} first elucidated the relationship between narrow/medium-band photometric colors and stellar metallicity. As of today, the large-sample spectroscopic surveys, such as SDSS/SEGUE \citep{Yanny2009,Rockosi2022}, LAMOST, and APOGEE,  provide excellent calibration labels for this method. \citet{2015ApJ...799..134Y} obtained the typical relation between the width of stellar loci and the range of metallicity by analyzing spectroscopic stellar parameters and SDSS DR9 \citep{2012ApJS..203...21A} photometric data of Stripe 82. Following this,  
\citet{2015ApJ...803...13Y} developed a method, based on fitting relationships between metallicity and photometric colors, and used this relationship to predict metallicity for stars with photometric colors but lacking spectroscopy. By adopting a similar method, \citet{2022ApJ...925..164H} obtained atmospheric-parameter estimates for over 20 million stars using SMSS DR2 photometic data and LAMOST DR9 and APOGEE DR17 spectroscopic data. \citet{2023ApJ...957...65H} obtained stellar parameters for more than 20 million additional stars in the Northern sky from the SAGES survey \citep{2023ApJS..268....9F}. We here will adopt the same technique to derive metallicity, effective temperature, and surface gravity for KIC stars from the KIS and synthesized Str{\"o}mgren photometry from Gaia XP spectra.

\subsection{Metallicity}\label{subsec:Metallicity}
\begin{figure}[ht!]
\epsscale{1.2}
\plotone{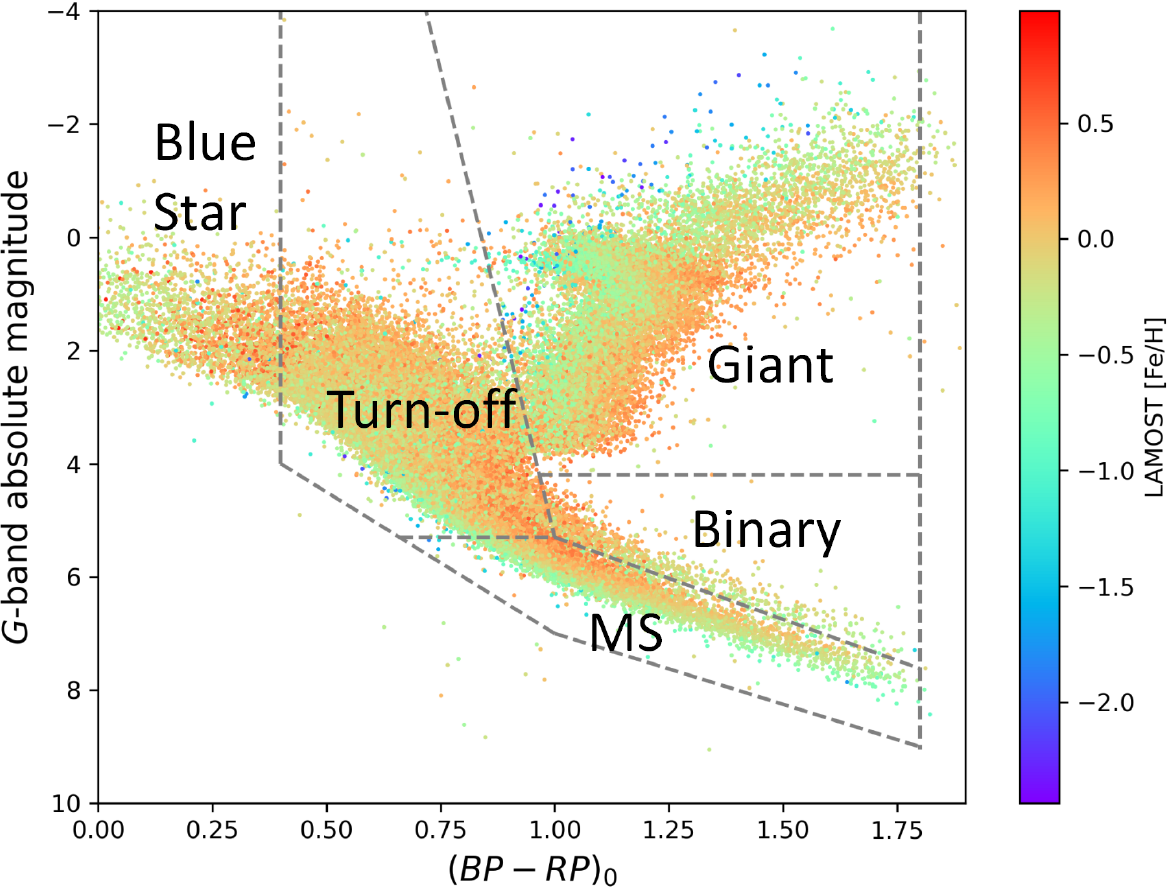}
\caption{Color–absolute magnitude diagram of the training sample defined in Section 4.1, coded by LAMOST metallicity, as shown in color bar to the right. The dashed lines represent the cuts to separate different types of stars, i.e., the main-sequence (MS) stars, binary stars, giant stars, turn-off stars, and blue stars with $(BP - RP)_0 < 0.40$. \label{fig:division}}
\end{figure}

To derive photometric metallicity, we use stars in common with the LAMOST-KIC dataset as our training set. In total, there are over 77,000 KIC stars observed by LAMOST with $g$-band SNR $\ge 20$.
Using these training stars, we aim to establish relationships between spectroscopic metallicity and stellar colors taken from either the KIS or Str{\"o}mgren photometry synthesized from Gaia XP spectra. Generally, the relationships are trained separately for dwarf and giant stars. To achieve better precision, we here train the relationships across five luminosity classes, including giants with $(BP - RP)_0 < 1.8$, main-sequence stars with $(BP - RP)_0 < 1.8$, binaries, turn-off stars, and blue stars with $(BP - RP)_0 <0.4$, based on their positions in the color-absolute magnitude diagram (see Figure~\ref{fig:division}).
We note that the cuts to select these classes are empirically determined.

To show the sensitivities of KIS $U$ and  Str{\"o}mgren m$_1 \equiv(v-b) - (b-y)$ on metallicity, two examples are shown in Figure~\ref{fig:stellarloci} for main-sequence and giant stars, respectively.
The plots clearly demonstrate that sequences at different metallicities ranging from around [Fe/H] = $-2.0$ to [Fe/H] = $+0.5$, as both colors $(U-BP)_0$ and [m$_1$]$\equiv(v-b)_0 - (b-y)_0$ change with the $BP$ for typical FGK-type stars\footnote{Here the reddening coefficients of the $U$ and Str{\"o}mgren bands are all taken from \url{http://svo2.cab.inta-csic.es/theory/fps/}.}.
Instead of using two-dimensional polynomial functions, as employed by previous studies \citep[e.g.,][]{2015ApJ...803...13Y, 2022ApJ...925..164H}, we adopt the random forest machine-learning method to model the relations ${\rm [Fe/H]} = f((U-BP)_0, (BP-RP)_0)$ and ${\rm [Fe/H]} = f([m_1], (BP-RP)_0)$ separately for the five luminosity classes.

\begin{figure}[ht!]
\epsscale{1.2}
\plotone{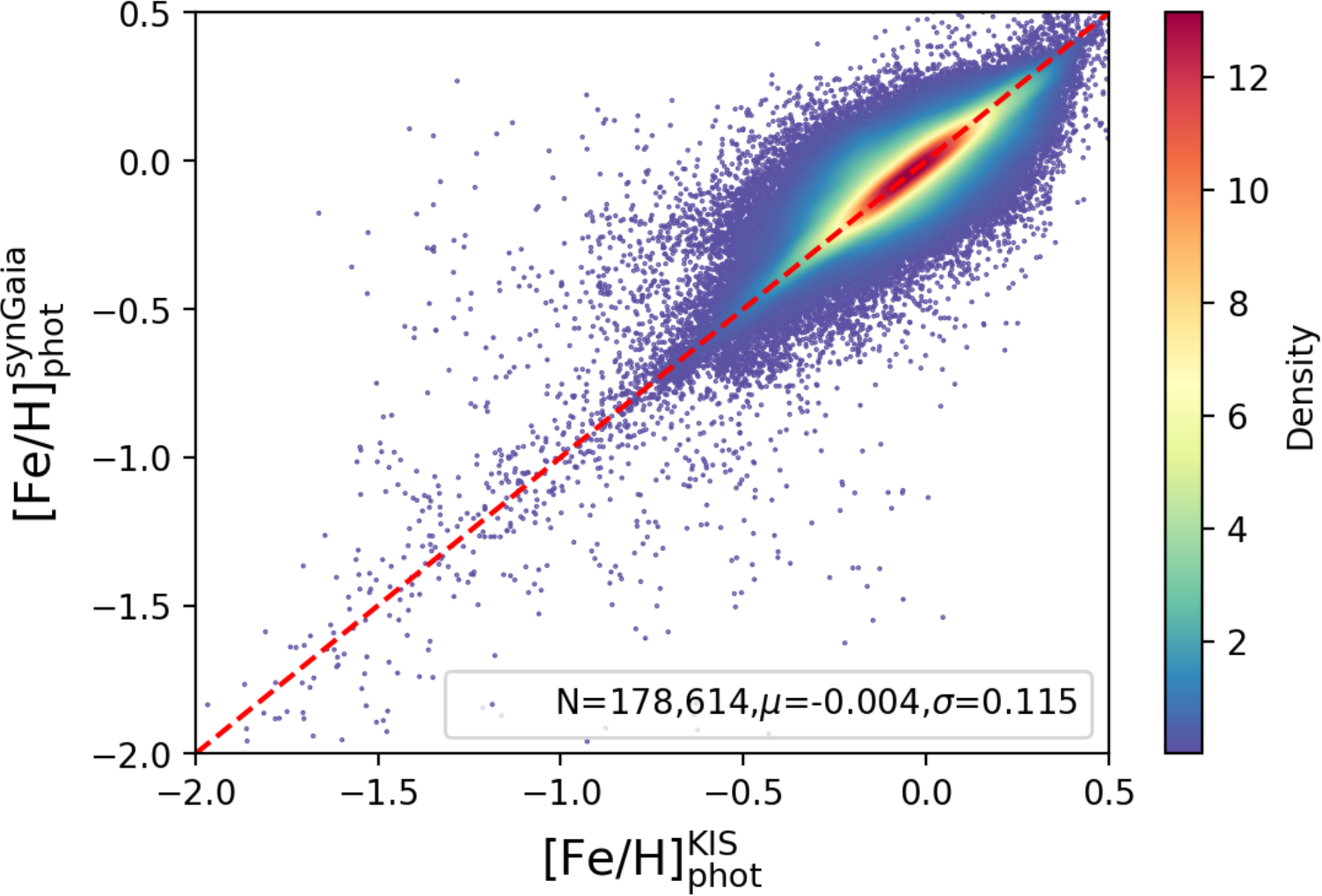}
\caption{Comparison of photometric-metallicity estimates derived from KIS photometry with those having synthesized Str{\"o}mgren from Gaia XP spectra. The red-dashed line is the one-to-one line. The color bar at right codes the number density of stars. The numbers of stars, mean offset, and dispersion are provided at the bottom right. \label{fig:twomethod}}
\end{figure}

After establishing the metallicity-color relations through training, we applied them to the entire KIC stellar sample to derive their photometric metallicities.
In total, we derived photometric metallicities for 179,413 KIC stars using the model trained with KIS photometry, and for 189,727 stars using the model trained on Gaia XP spectral-synthesis-generated photometric data. Overall, this yields metallicity estimates for 191,551 stars, representing 95\% of the total KIC stellar sample.
First, as an internal check, the metallicity estimated from KIS is compared to that derived from the synthesized  Str{\"o}mgren photometry in Figure~\ref{fig:twomethod}.
No offset is found between the estimates from the two relations, with a minimal scatter of only 0.12\,dex. This suggests an intrinsic precision of 0.08\,dex, assuming equal contributions to the scatter from both relations.

\begin{figure*}[ht!]
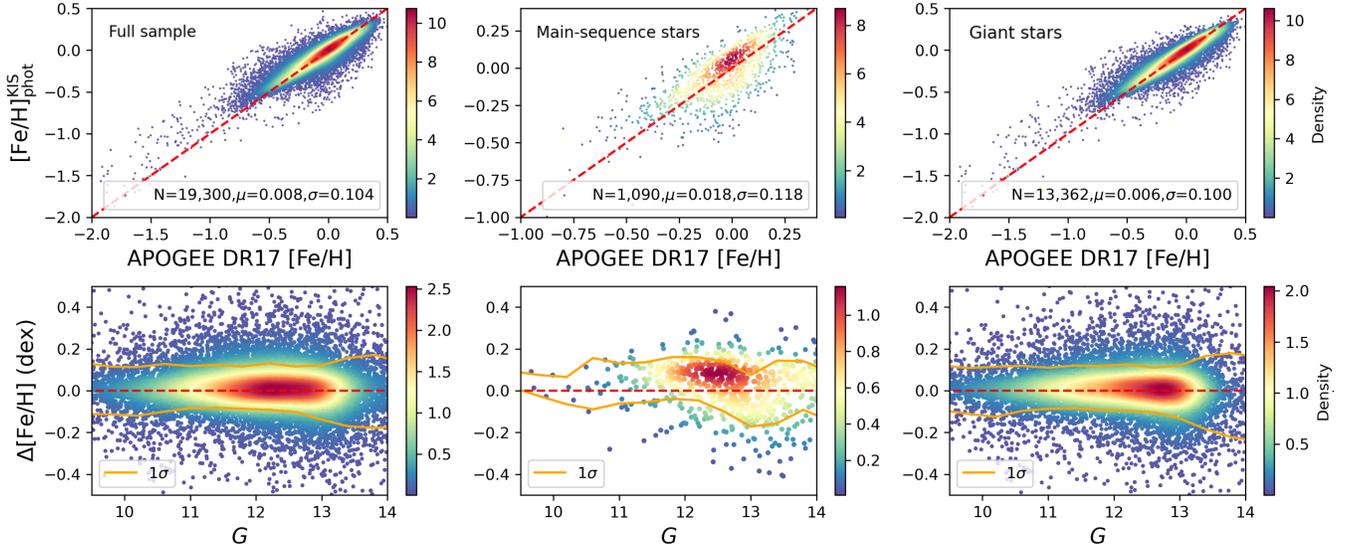

\gridline{\fig{RFfit/feh_kis.png}{1\textwidth}{}
          }
\caption{Upper panels: Comparison of photometric metallicity estimates from the KIS photometry with those from APOGEE DR17 for the full sample (left panel), main-sequence stars (middle panels) and giant stars (right panels).
The red-dashed lines are the one-to-one lines. The numbers of stars, mean offset, and dispersion are provided at the lower right.  Lower panels: Residuals of the metallicity differences (${\rm [Fe/H]_{~phot}}-{\rm [Fe/H]_{APOGEE}}$), as a function of Gaia $G$ magnitude. The red-dashed line is the zero level. The golden lines represent the $1\sigma$ scatter. The color bar at the right of each panel codes the number density of stars. 
\label{fig:KISFeHcomparison}}
\end{figure*}

To check the accuracy of the derived photometric metallicity, we cross-matched and compared KIC with results from APOGEE DR17. Overall, the photometric metallicities estimated from KIS and Str{\"o}mgren colors exhibit excellent agreement with those from APOGEE DR17, with negligible offsets and a very small scatter of around 0.10\,dex (see Figures \ref{fig:KISFeHcomparison} and \ref{fig:GaiaGenFeHcomparison}). 
However, we find that the photometric metallicities are slightly higher than those from APOGEE DR17, likely due to differences in the metallicity scales between LAMOST and APOGEE \citep{2024arXiv240802171H}.
This consistency holds across all stellar types, except for blue stars, where the limited number of KIC-APOGEE common stars prevents a meaningful comparison (see Figures \ref{fig:KISFeHcomparison}, \ref{fig:GaiaGenFeHcomparison}, \ref{fig:turnoffkis}, and \ref{fig:turnoffgaia}).
The precision for  main-sequence, turn-off, giant, and binary stars are 0.12, 0.10, 0.10, and 0.18\,dex, respectively.

\begin{figure*}[ht!]
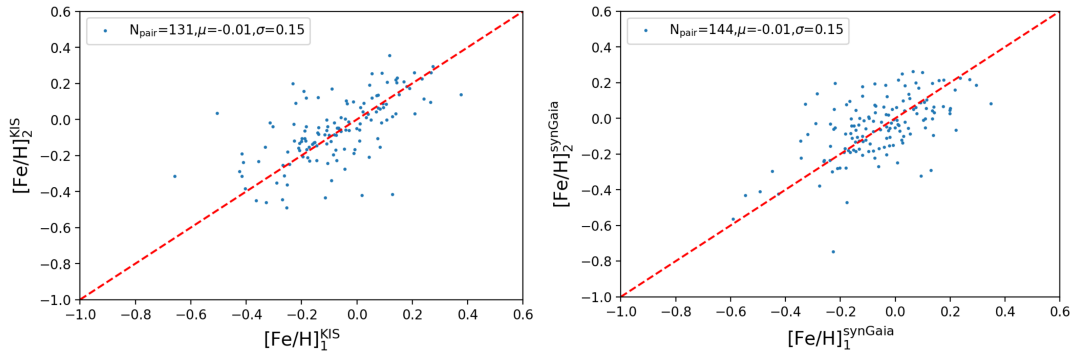

\gridline{\fig{RFfit/ultra_wide_binary.png}{0.8\textwidth}{}}
\caption{Comparison of photometric [Fe/H] estimates between wide binary members. Left panel: Comparison of our method based on synthetic photometric data generated from Gaia XP spectra. Right panel: Our method based on KIS photometric data. The red-dashed lines are the one-to-one lines. The numbers of stars, mean offset, and dispersion are provided at the upper left of each panel. 
\label{fig:widebinary}}
\end{figure*}

To further evaluate the accuracy of the photometric metallicities, our sample is cross-matched with wide binaries selected from Gaia DR2 \citep{2020ApJS..246....4T}, which are expected to have identical metallicities due to their identical birthplace and formation time.
In total, 131 and 144 wide binaries are found to have photometric estimates of metallicity measured from KIS and Str{\"o}mgren colors, respectively. As shown in Figure \ref{fig:widebinary},  the offsets are within 
0.02\,dex, with a scatter of approximately 0.15\,dex, demonstrating the consistency of metallicities between stars in the same binary system.

\begin{figure*}[ht!]
\gridline{\fig{RFfit/stellar_loci_feh.png}{0.8\textwidth}{}
          }
\caption{Distributions of the training-sample main-sequence stars (left column of panels) and giant stars (right column of panels) in the $(U - BP)_0$ versus  $(BP - RP)_0$ plane (top column) and the [m$_1$] versus $(BP - RP)_0$ plane (right column), coded by LAMOST metallicity ([Fe/H]), as shown by the color bars to the right of each panel. The dashed lines represent equal-metallicity sequences ranging from $+0.5$ (up) to $-1.0$ (bottom panel) for dwarf stars and $+0.5$ (up) to $-2.0$ (bottom panel) for giant stars, respectively. 
These sequences are obtained by fitting second-order polynomials to stars in the metallicity range $|$[Fe/H]$-$[Fe/H]$_i| < 0.1$, where [Fe/H]$_i$ represent the marked metallicity of these sequences. The numbers of stars in each panel are provided at the lower right of the panel.
\label{fig:stellarloci}}
\end{figure*}

\begin{figure*}[ht!]
\gridline{\fig{RFfit/feh_gaiagen.png}{1\textwidth}{}
         }
\caption{Similar to Figure~\ref{fig:KISFeHcomparison}, but for photometric-metallicity estimates from  Str{\"o}mgren photometry synthesized from the Gaia XP spectra.
\label{fig:GaiaGenFeHcomparison}}
\end{figure*}

\subsection{Effective Temperature}\label{subsec:teff}

In the previous subsection, we discussed the methodology for estimating the photometric metallcity for the great majority of the KIC stars. These photometric metallicities will serve as inputs to the training sets for subsequent steps in this  work.  First, we consider the effective temperature,  $T_{\rm eff}$.

\begin{figure*}[ht!]
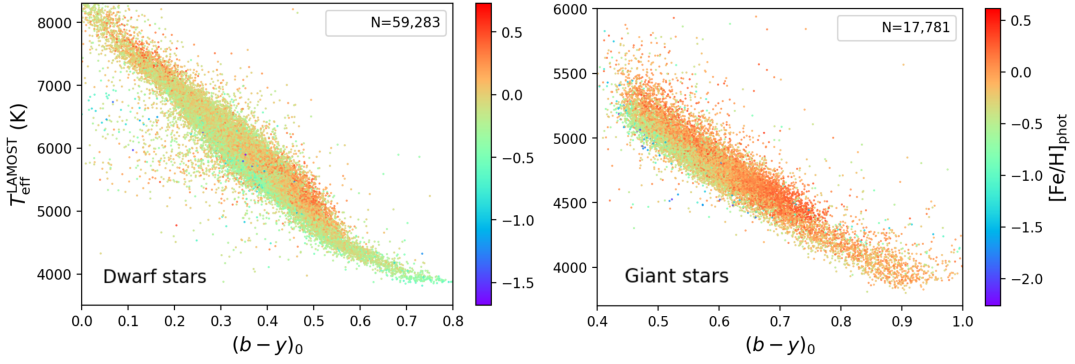

\gridline{\fig{RFfit/stellar_loci_teff.png}{0.8\textwidth}{}
          }
\caption{The relations between $T_{\rm eff}$ and color for dwarf stars (including main-sequence stars, turn-off stars and blue stars, shown in left panel) and giants (right panel). The $T_{\rm eff}$ data is adopted from LAMOST DR10, and the [Fe/H] data is from our photometric estimates.  The numbers of stars are shown in the upper right of each panel. The color bar to the right of each panel codes the photometric-metallicity estimates.}
\label{fig:teffloci}
\end{figure*}

Figure \ref{fig:teffloci} shows the $T_{\rm eff}$ versus $(b-y)$ color plots for giants and dwarfs (hereafter dwarfs represent main-sequence stars, turn-off stars, and blue stars), respectively. In this case, the photometry is synthetically generated from the Gaia XP spectra, the effective temperatures are taken from the LAMOST DR10 spectroscopic data, and the metallicities used to color-code the legend are those obtained from the photometric fits described in Section \ref{subsec:Metallicity}. From inspection, similar to the color-color map, the stars are on distinct loci and are stratified because of their different metallicities. We obtain estimates of effective temperature with the following relationship:

 \begin{equation}
T_{{\rm eff},i}=f((b-y)_0,{\rm [Fe/H]}).
\end{equation}

Again, a random forest regressor machine-learning method is adopted to train the 
$T_{\rm eff}$-color-[Fe/H] relation. For the training set, the spectroscopic effective temperature is chosen from the LAMOST DR10 low-resolution spectra (LRS) AFGK catalog, with SNR$_g > 20$. 
The total training set consists of 71,892 stars. In this section, we use a different division of stars compared to Section \ref{subsec:Metallicity} to train the $T_{\rm eff}$-color-[Fe/H] relation separately. Here, the stars are divided into three types, dwarfs, binaries, and giants, rather than the five types used in Section \ref{subsec:Metallicity}. After comparing the accuracy of the metallicities derived in Section~\ref{subsec:Metallicity} with the APOGEE DR17 data, we adopted the photometric metallicity estimates from the method with higher accuracy for each type of star.  For stars classed as dwarfs, the photometric metallicity estimates trained by KIS and Gaia photometry were adopted. For stars classed as binary or giant, we employed the photometric metallicities trained by synthetic photometry data from the Gaia XP spectra.

\begin{figure*}[ht!]
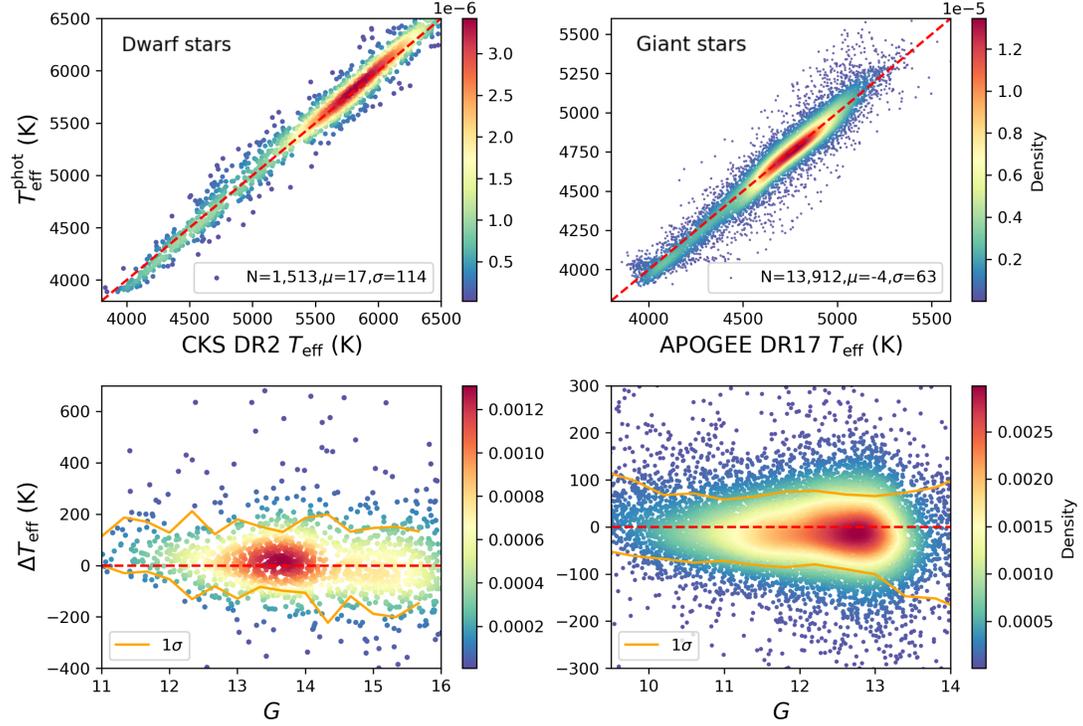

\gridline{\fig{/RFfit/teff.png}{0.8\textwidth}{}
          }

\caption{Comparison of $T_{\rm eff}$ obtained by our method with that from CKS DR2 (left column of panels for dwarf stars) and APOGEE DR17 (right column of panels for giant stars).  The red-dashed lines in the upper panels are the one-to-one lines. The numbers of stars, mean offset, and dispersion are provided at the lower right of each upper panel.  The lower panels show the residuals in the sense of $T_{\rm eff}^{\rm phot}-T_{\rm eff}^{\rm APOGEE}$, as a function of $G$ magnitude. The red-dashed lines in the lower panels represent the zero level.  The golden lines in the lower panels represents the $1\sigma$ scatter. The color bar to the right of each panel codes the number density of stars in the figure.
\label{fig:teffcomparison}}
\end{figure*}

We then applied the trained relations to all three types of stars, resulting in effective temperature estimates for a total of 189,727 KIC stars. Figure \ref{fig:teffcomparison} shows a comparison with the spectroscopic effective temperatures from CKS DR2 \citep{2017AJ....154..107P, 2018AJ....155...89P} and APOGEE DR17. 
For the giant sample, there is a tiny offset of only +4\,K and a dispersion of 63\,K, when analysing the  difference of our photometric $T_{\rm eff}$ minus APOGEE DR17. For the Binaries, the offset is +110\,K with 
a scatter of 247\,K. For the dwarf sample, we compared our photometric $T_{\rm eff}$ estimates to those from CKS DR2, rather than APOGEE DR17, as the latter's pipeline is primarily designed for giant stars.
The results show a small offset of +17\,K (this work minus CKS) with a scatter of 114\,K.

\subsection{Surface Gravity}\label{subsec:logg}

\begin{figure*}[ht!]
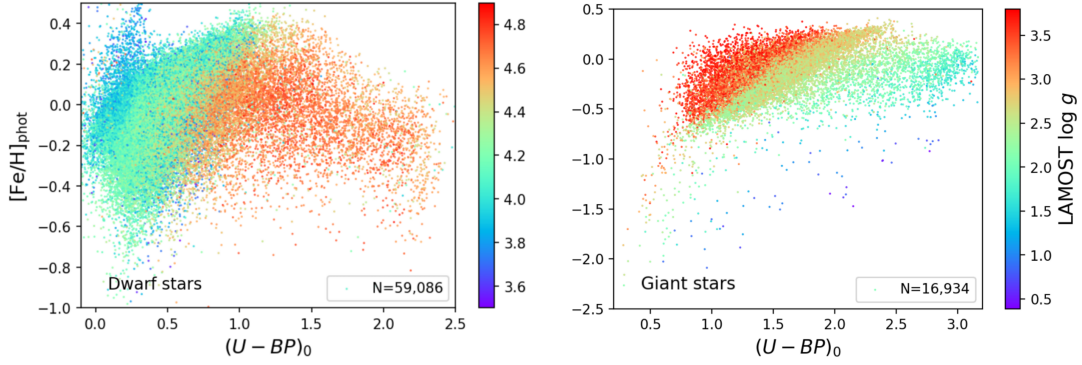

\gridline{\fig{RFfit/stellar_loci_logg.png}{0.8\textwidth}{}
          }
\caption{The relations between log $g-(U-BP)_{0}$ and photometric-metallicity estimates for main-sequence stars (left panel) and giant stars (right panel). The log $g$ data is from LAMOST DR10 and the metallicities are obtained by our photometric estimates.  The numbers of stars are provided in the lower right of each panel. The color bar to the right of each panel codes the LAMOST DR10 estimate of log $g$. 
\label{fig:loggloci}}
\end{figure*}

We now consider surface gravity estimates for the KIC stars based on the photometric color and photometric metallicities. Figure \ref{fig:loggloci} shows the log $g$-$(U-BP)$ color plots for giant and dwarf  stars. In this case, the photometry data is from KIS DR2 and Gaia DR3. The metallcities are from the result of photometric estimates as described in Section \ref{subsec:Metallicity}, and the surface gravities used for training are from the LAMOST DR10 spectroscopic data. The $U$-band photometry contains information about both the Balmer jump (which correlates with surface gravity) and metallicity.
With the photometric metallicity fixed, the color $(U-BP)_0$ can be further used to constrain log~$g$.
As seen in the plots, the stars are stratified due to their different values of log $g$. Once again, we employed a random forest regressor machine-learning method to train the log $g$-color-[Fe/H] relations for different types of stars, following the same technical treatments as used for effective temperature (see Section \ref{subsec:teff}). The total training set consists of 68,063 stars. From Figure \ref{fig:loggloci}, we note that, among the dwarf stars, some with low log $g$ values, located at $(U-BP)_0 \sim 0.5$ and ${\rm [Fe/H]} \sim -0.6$, do not conform to the overall log $g$ gradient changes in the $(U-BP)_0$–$(BP-RP)_0$ diagram. Upon further examination, we found that these stars are located at the boundary between turn-off stars and sub-giant stars. Therefore, the discrepancy for these stars is possibly due to their stellar classification; they are better classified as giants rather than dwarfs.
We then applied the trained relations to all types of stars, obtaining surface gravity estimates for 189,727 KIC stars. Figure \ref{fig:loggcompare} shows a comparison between our method and the spectroscopic surface gravity estimates from CKS DR2 (for dwarf stars) and APOGEE DR17 (for giants and binaries). The result for dwarfs exhibits an offset of $-$0.01\,dex (this work minus CKS DR2) and a dispersion of 0.14\,dex. As for result of giant stars compared with APOGEE DR17, the offset is +0.17\,dex (this work minus APOGEE DR17) and the dispersion is 0.19\,dex. For binaries, the offset is +0.03\,dex (this work minus APOGEE DR17) and the dispersion is 0.09\,dex. The moderate offset in surface gravity for giant stars is primarily due to the scale difference between LAMOST (used as the training set) and APOGEE. By comparing over ten thousand common stars between LAMOST and APOGEE, we detected a similar offset of approximately 0.14 dex in surface gravity for giant stars.

\begin{figure*}[ht!]
\gridline{\fig{/RFfit/logg.png}{0.8\textwidth}{}
          }
\caption{Comparision of log $g$ estimates obtained by our method with CKS DR2 (left column of panels for dwarf stars) and APOGEE DR17 (right column of panels for giant stars). The red-dashed lines in the upper panels are the one-to-one lines. The numbers of stars, mean offset, and dispersion are provided at the lower right of each panel. The lower panels show the residuals in the sense of  $\log g_{\rm phot} - \log g_{\rm APOGEE/CKS}$, as a function of $G$ magnitude.
The red-dashed lines in the lower panels represent the zero level.  The golden lines in the lower panels represents the $1\sigma$ scatter. The color bar to the right of each panel codes the number density of stars in the panel.
\label{fig:loggcompare}}
\end{figure*}


\subsection{Uncertainty Analysis}\label{subsec:uncertainty}

Random forest machine-learning methods generally do not provide uncertainty estimates for the derived parameters of individual stars. To address this, we estimate the uncertainty of each parameter using the Monte Carlo (MC) method. The MC simulation accounts for both the uncertainties in the input quantities and the mapping relations defined by the random forest regressor.

Using [Fe/H] as an example, we train the color–[Fe/H] relations with the random forest algorithm 1000 times. In each iteration, we sample the photometric errors from KIS/Gaia-synthesized photometry and Gaia data, as well as the uncertainties in extinction. All these errors are assumed to follow a Gaussian distribution. 
We then apply each relation to all KIC stars, again sampling their photometric and extinction uncertainties under the assumption of Gaussian distributions. For each star, this process yields a distribution of photometric [Fe/H], with the dispersion serving as the uncertainty. 
Figure \ref{fig:error} shows an example of the final distribution of 1000 simulated photometric [Fe/H] estimates and dispersion.
Following the same approach, we derive the uncertainties for photometric $T_{\rm eff}$ and log~$g$. 

To validate the reliability of the uncertainty estimates, we compare our results with those from APOGEE DR17. First, we divide the stars into magnitude bins within the specified magnitude range. For each bin, we calculate the dispersion of the differences between our derived [Fe/H] and those from APOGEE DR17, treating this dispersion as the reference uncertainty for that group. Then we compare these reference uncertainties with the mean uncertainties obtained from our method. As shown in 
the right panel of Figure \ref{fig:error}, the uncertainties from the MC simulations are in excellent agreement with those with APOGEE DR17 uncertainties, with an offset of 0.01 dex and the dispersion is 0.04 dex, confirming the robustness of the uncertainty calculations.

\begin{figure*}[ht!]
\gridline{\fig{uncertainty/error.png}{0.8\textwidth}{}}
\caption{Left panel: An example of the distribution of final photometric estimates of [Fe/H] yielded by the Monte Carlo simulations. The median value of this distribution is marked by a dashed-green line and the estimate by LAMOST is marked by a red-dashed line.
Right panel: Comparison of the mean uncertainties derived from our Monte Carlo method with those obtained from APOGEE DR17.
The `STD' is estimated by calculating the dispersion of the metallicity difference between photometric method and APOGEE DR17 across various magnitude bins. We use 100 bins, evenly spaced between magnitudes 10 and 16.  The red-dashed line is the one-to-one line. \label{fig:error}}
\end{figure*}

\begin{figure*}[ht!]
\epsscale{0.9}
\plotone{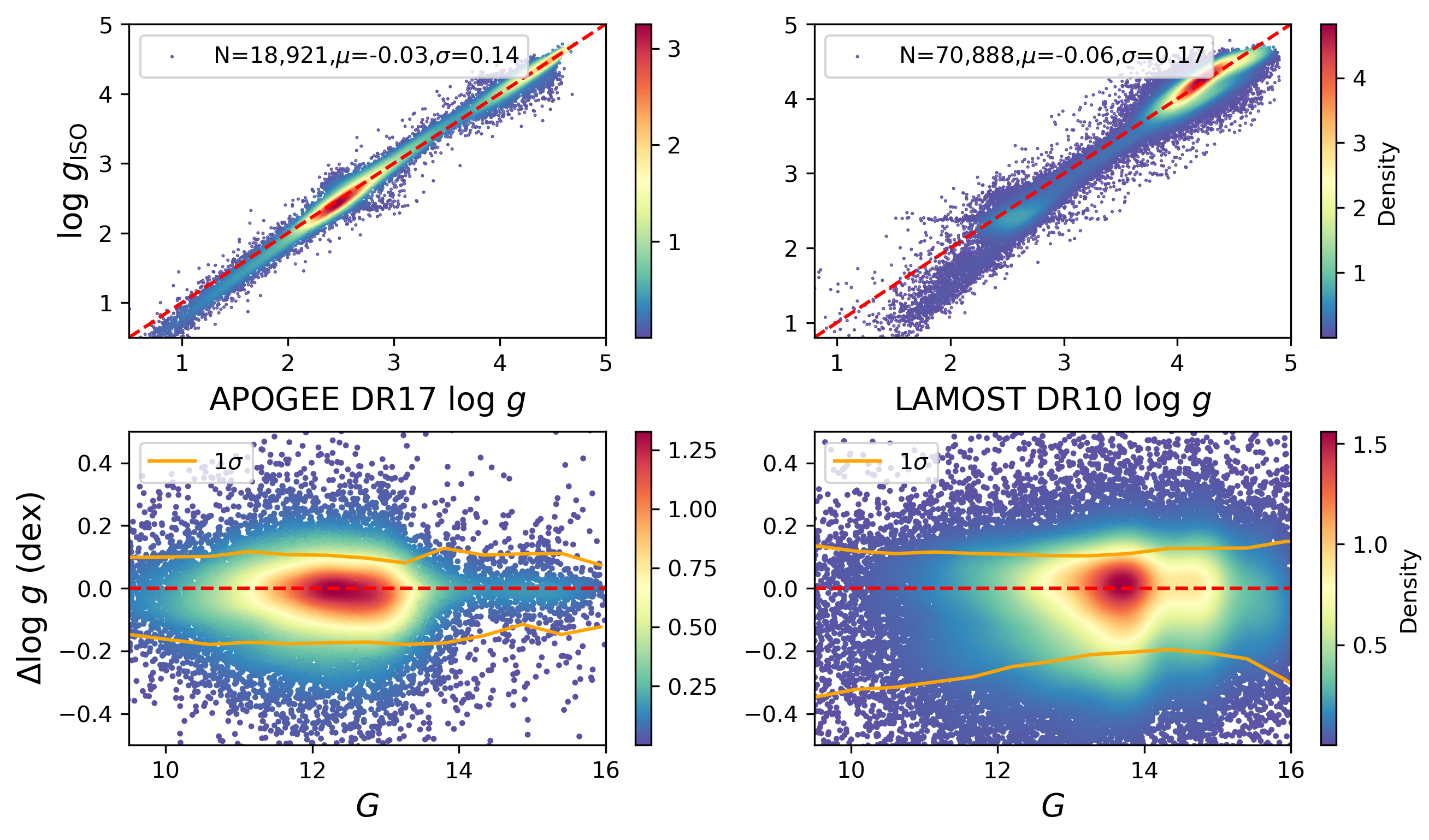}
\caption{Comparision of log $g$ obtained by the isochrone-fitting method with APOGEE DR17 (left column of panels) and LAMOST DR10 (right column of panels).
The red-dashed lines in the upper panels are the one-to-one lines. The numbers of stars, mean offset, and dispersion are provided at the upper left of each upper panel.  
The lower panels show the residuals in the sense of $\log g_{\rm ~isochrone~fitting} - \log g_{\rm APOGEE/LAMOST}$, as a function of $G$ magnitude.  The red-dashed lines in the lower panels represent the zero level.  The golden lines in the lower panels represents the $1\sigma$ scatter.  The color bar to the right of each panel codes the number density of stars in the panel.
\label{fig:loggISOcompare}}
\end{figure*}

\section{Stellar Age Estimation}\label{subsec:Isofit}

\subsection{Bayesian Estimate}
In this section, we derive ages, masses, and radii of the KIC stars from isochrone fitting based on a Bayesian approach. The methods we apply are similar to those used by \citet{2005A&A...436..127J} and \citet{2022ApJ...925..164H}. Essentially, we match the observed parameters with the theoretical results given by stellar-evolution models, and obtain these estimates from the models. 

The observed parameters we employ are: 1) the intrinsic color $(BP-RP)_0$ and $G$-band absolute magnitude of the KIC stars, corrected by our 
3-D extinction map; 2) the stellar metallicities. 
For metallicity, we combined the spectroscopic metallicities from APOGEE and LAMOST, where available, along with the photometric metallicities obtained by our methods. For metallicities from different sources, we used the following criteria. For a given KIC star, if the APOGEE data has a metallicity that is obtained from a spectrum with SNR larger than 30, this metallicity is chosen. If this is not available, and there is an available metallicity obtained from a LAMOST LRS spectrum whose $g$-band SNR is greater than 30, we select that estimate. If neither of these are available,  we select the photometric metallicity using the same strategy employed for selecting data for the training set for models used to fit the $T_{\rm eff}$-color-[Fe/H] and log $g$-color-[Fe/H] relations, according to the assigned object type of the star. 

For the stellar-evolution models, we used the PARSEC isochrones \citep{2012MNRAS.427..127B}. For the ages and metallicities of the models, we divided the grid over the age ranging from 0.1 to 15.2 Gyr and [M/H] from $-2.2$ to +0.5. The step of the age grid is 0.2 Gyr for models with ages younger than 1.2 Gyr, and 0.5 Gyr for models whose ages are older than 1.2 Gyr. The step in [M/H] is 0.02\,dex. This yields a grid of $1.38\times10^6$ stellar model points.

There remains the problem that the theoretical metallicities given by PARSEC are in the form of [M/H], but the observed metallicities are in the form of [Fe/H]. We transformed the metallicities measured by [Fe/H] to [M/H] using  Equation 6 from \citet{2005essp.book.....S}. 
 \begin{equation}
{\rm [M/H]} =  {\rm [Fe/H]}+\log(0.694+10^{{\rm [\alpha/Fe]}}+0.306).
\end{equation}
For  [$\alpha$/Fe] in this relation, we fitted the [$\alpha$/Fe]-[Fe/H] relation from the APOGEE [$\alpha$/H] and [Fe/H]. The fitting relation is a sixth-order polynomial; the parameters for this polynomial are listed in Table \ref{tab:MHFeH}.

For the Bayesian estimation method, there are three parameters that decide stellar evolution: age $\tau$, mass m, and metallicity $Z$. Thus, the posterior probability distribution function of the stellar parameters can be described as: 

 \begin{equation}
f(\tau,M,Z)\propto f_0(\tau,M, Z)\mathcal{L}(\tau,M,Z)\text{,}
\end{equation}

\noindent where $f_0$ is the prior distribution of the parameters. In this work, we assumed that age and metallicity [M/H] follow a uniform distribution. For the mass, we assumed that it follows a power-law distribution given by \citet{1955ApJ...121..161S}:

 \begin{equation}
f_0(M)\propto M^{-2.35}\text{.}
\end{equation}

The prior distributions are independent of each other. $\mathcal{L}$ is the likelihood function of the parameters, which can be described as:

 \begin{equation}
\mathcal{L}(\tau, M, Z)=\mathop{\Pi}\limits_{i=1}^{n}(\frac{1}{\sqrt{2\pi}\sigma_i}) \times e^{-\frac{\chi^2}{2}}\text{.}
\end{equation}

Here $\chi^2$ is defined as: 
 \begin{equation}
\chi^2=\sum_{i=1}^n(\frac{O_i-T_i(\tau,M,Z)}{\sigma_i})^2\text{,}
\end{equation}

\noindent where the $O$ represents observational parameters, including the 
$G$-band absolute magnitude, intrinsic $(BP-RP)_0$ color, and metallicity [M/H]. $T$  are the theoretical values of those parameters given by the isochrone model under a specific set of parameters for $\tau$, $M$, and $Z$.

With this procedure we can obtain the posterior probability distribution function (PDF), denoted as $\mathcal{L}(P|\tau, M, Z)$, for the parameter of interest.
The parameters to be determined include stellar mass, age, surface gravity, and radius.
For each parameter, we then calculate the PDF for each star using our Bayesian approach.
The final estimate of each parameter for a given star is taken as the median of the resulting posterior PDF, with its uncertainty defined as half the difference between the 84th and 16th percentile values of the posterior PDF.
The estimated physical parameters are then compared with independent measurements to assess their accuracy.

\subsection{Comparison with APOGEE and LAMOST}\label{subsec:LA}
The resulting log~$g$ values are compared with those from APOGEE DR17 and LAMOST DR10.
As shown in Figure~\ref{fig:loggISOcompare}, the values from isochrone fitting are consistent with the spectroscopic measurements. The mean offsets are only $-0.03$\,dex (isochrone fitting minus APOGEE) and (isochrone fitting minus APOGEE) and $-0.06$\,dex (isochrone fitting minus LAMOST), with small scatters of 0.14 and 0.17\,dex, respectively.
These comparisons indicate that our log~$g$ estimates from isochrone fitting are more accurate than those derived from stellar colors, as described above.

\subsection{Comparison with SD18}\label{subsec:SD18}

To validate our age and mass estimates, we cross-matched our results with those from \citet[][hereafter SD18]{2018MNRAS.481.4093S}, which provides a catalog of stellar ages and masses for approximately 3 million stars, derived from spectroscopic data from existing surveys combined with Gaia parallax measurements.

\begin{figure*}[ht!]
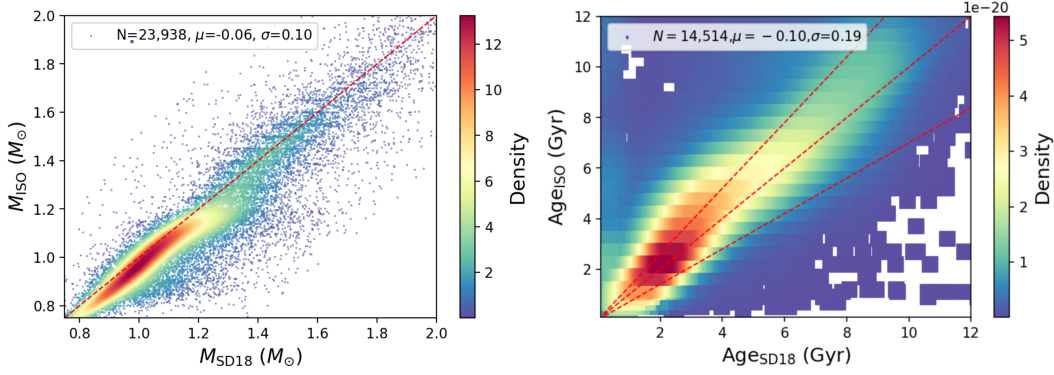

\gridline{\fig{isofit/age_sd18.png}{0.8\textwidth}{}
          }
\caption{Comparisons of stellar mass (left panel) and age (right panel) estimates between this work and \citet[][hereafter SD18]{2018MNRAS.481.4093S} for, respectively, nearly 24,000 main-sequence  stars and 14,000 main-sequence turn-off stars in common. The red-dashed ine in the left panel is the one-to-one line.  In the right panel, the red-dashed lines indicate Age$_{\rm ISO} = 1.3$Age$_{\rm SD18}$ and Age$_{\rm ISO} = 0.7$Age$_{\rm SD18}$, respectively. The numbers of stars, mean offset, and dispersion are provided in the upper right of each panel. The color bar to the right of the each panel codes the density number of stars in the panel. }
\label{fig:ageSD18}
\end{figure*}

\begin{figure*}[ht!]
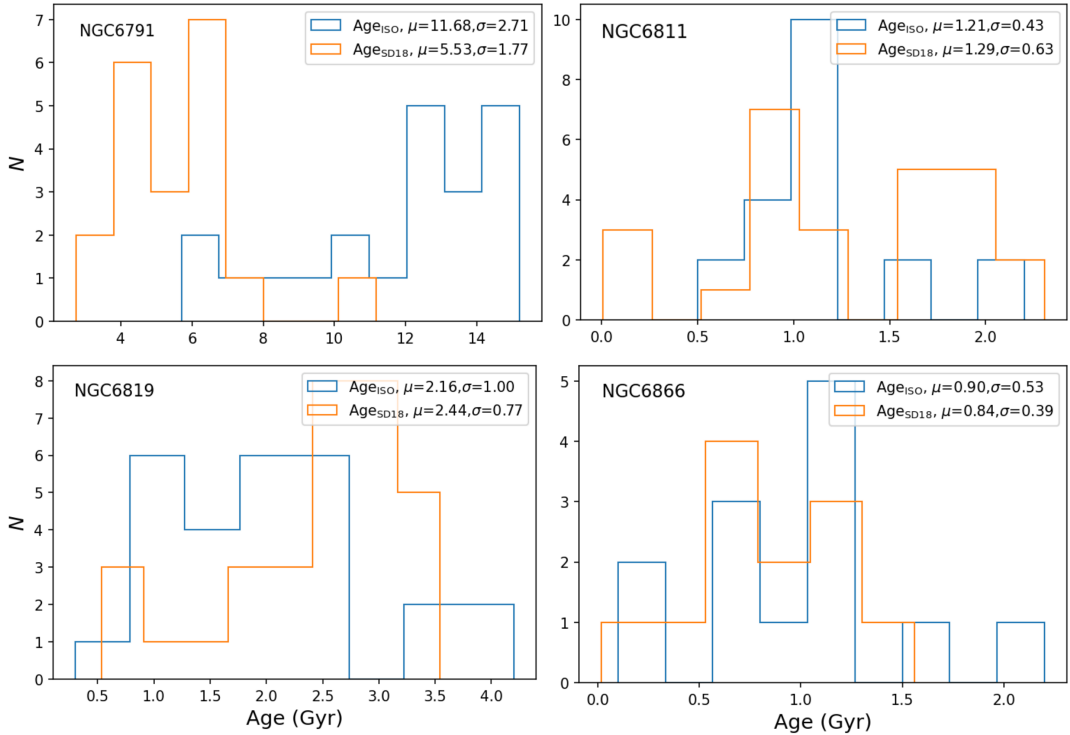

\gridline{\fig{/isofit/age_cluster.png}{0.8\textwidth}{}
          }
\caption{Age distributions for member stars of the four open clusters (NGC 6791, NGC 6811, NGC 6819, and NGC 6866) in the Kepler field. The blue lines represent the ages derived in this work, while the orange lines indicate the ages from SD18. The mean and dispersion of these distributions are marked in top-right corner of each panel.}
\label{fig:ageCluster}
\end{figure*}

First, the stellar masses are compared to those from SD18 in Figure \ref{fig:ageSD18}.
Overall, the consistency is very good, with a mean offset of $-0.06$~$M_{\odot}$  (our values minus those of SD18) and a scatter of $0.10$~$M_{\odot}$.
Note that giant stars are excluded from this comparison, as their mass estimates are highly sensitive to the uncertain parameter of mass loss, for which we adopted a constant value of $\eta_{\rm Reimers} = 0.2$, following the recommendation in PARSEC. We will later assess the masses of giant stars using asteroseismic estimates.
Secondly, we compare our stellar ages with those from SD18, as shown in Figure \ref{fig:ageSD18}.
Only turn-off stars are included in this comparison, as their ages can be reliably constrained through isochrone fitting. The relative age ratio $(\tau_{\text{ISO}}-\tau_{\text{SD18}})/(\tau_{\text{ISO}})$ shows a mean offset of +10\%, with a dispersion around 19\%.
To further evaluate the precision of isochrone-derived ages, we selected members of open clusters using the method described in Section~\ref{subsec:extinction}. Figure~\ref{fig:ageCluster} shows the age distributions for member stars of four open clusters in the Kepler field. The median values of these distributions are close to those reported by other independent studies \citep[][see Table \ref{tab:openCluster}]{2016AJ....151...66B,2016ApJ...831...11S,2019AA...623A.108B,2021AA...649A.178B}.

\begin{deluxetable}{ccccccc}
\tablenum{2}
\tablecaption{Polynomial-fitting Coefficients for the Relationship Between [$\alpha$/Fe] and [Fe/H]\label{tab:MHFeH}}
\tablewidth{0pt}
\tablehead{
\colhead{a$_0$} & \colhead{a$_1$} & \colhead{a$_2$} & \colhead{a$_3$} & \colhead{a$_4$} & \colhead{a$_5$} & \colhead{a$_6$}
}
\startdata
 $-$0.0676 & $-$0.3995 & $-$0.6846 & $-$0.1530 & 0.3205 & $-$0.1469 & 0.0355
\enddata
$[\alpha/\mathrm{Fe}]=a_0\times b^6+a_1\times b^5+a_2\times b^4+a_3\times b^3+a_4\times b^2+a_5\times b+a_6$\\
\tablecomments{b=[Fe/H]}
\end{deluxetable}

\subsection{Comparison with CKS}\label{subsec:CKS}
As previously described, the CKS is a high-resolution spectroscopic survey designed to determine the properties of exoplanets and their host stars in the Kepler field. Observations conducted with the Keck telescope have provided atmospheric parameters and other characteristics for approximately 1700 exoplanet-host stars. Using the Keck spectra, \citet{2017AJ....154..107P, 2018AJ....155...89P} derived the stellar atmospheric parameters (effective temperature, surface gravity, and metallicity) for the exoplanet-host stars. Based on these parameters, \citet{2017AJ....154..108J} further determined the masses and radii of these stars.

\begin{figure*}[ht!]
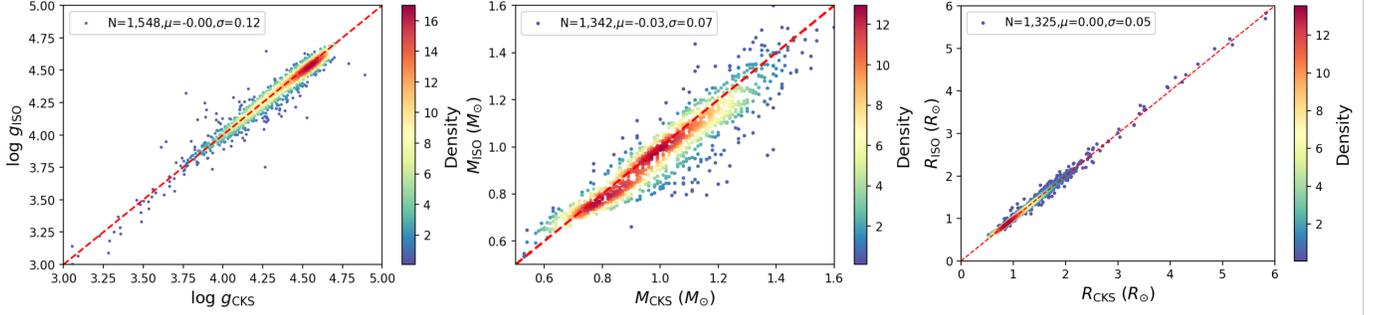

\gridline{                                                           
          \fig{/isofit/feh_m_r_cks.png}{1\textwidth}{}
          }
\caption{Comparisons of surface gravity (left panel), mass (middle panel) and radius (right panel) estimates between this work and the CKS survey. The numbers of stars, mean offset (this work minus CKS), and scatter are marked in the upper left of each panel. The red-dashed lines are one-to-one lines. The color bar to right of each panel codes the number density of stars in the panel.}
\label{fig:comparedCKS}
\end{figure*}

As shown in Figure \ref{fig:comparedCKS}, the isochrone-derived log~$g$ is in excellent agreement with that of CKS, with no offset and a minimal scatter of 0.12\,dex.
Both stellar mass and radius from isochrone fitting are also in very close agreement with CKS results. The offsets are negligible, with no offset for radius and 
only $-0.03$~$M_\odot$ (our values minus those of CKS) for mass.
The scatter is just 0.07~$M_\odot$ for mass and $0.05 R_{\odot}$ for radius.

\subsection{Comparison with APOKASC}\label{subsec:astroseismology}

Asteroseismology is an important technique in the field of stellar parameter measurement, as it enables precise estimates of mass, radius, and surface gravity. Here, we compare our results with those from the Apache Point Observatory Galactic Evolution Experiment and the Kepler Asteroseismic Science Consortium \citep[APOKASC;][]{2014ApJS..215...19P, 2018ApJS..239...32P}. The APOKASC catalog provides stellar parameters derived by combining asteroseismic data (such as frequency spacing $\Delta \mu$ and maximum oscillation frequency $\mu_{\rm max}$, from which mass and radius can be estimated) from the Kepler Asteroseismic Science Consortium (KASC) with spectroscopic data (such as $T_{\rm eff}$, [Fe/H]) from APOGEE.

\begin{figure*}[ht!]
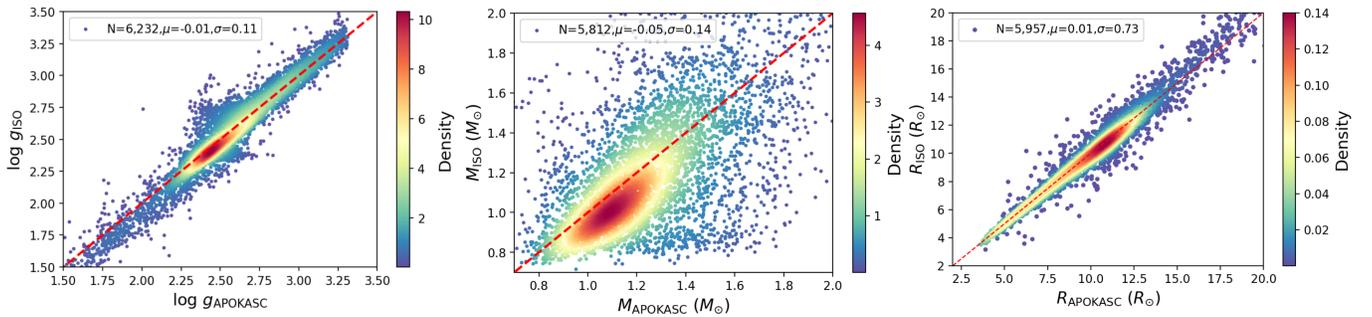

\gridline{                                                           
          \fig{/asteroseismology/apokasc.png}{1\textwidth}{}
          }
\caption{Similar to Figure~\ref{fig:comparedCKS}, but for comparison with APOKASC. \label{fig:compareapokasc}}
\end{figure*}

We compared our isochrone-derived log~$g$, mass, and radius with those from APOKASC. As shown in Figure~\ref{fig:compareapokasc}, all parameters estimated from isochrone fitting exhibit good agreement with APOKASC results. For log~$g$, the mean offset is only $-0.01$\,dex (our result minus APOKASC), with a dispersion of 0.11\,dex. For stellar radius, there is a offset of 0.01 $R_{\odot}$, and the scatter is only  $0.73 R_{\odot}$. For stellar mass, a slight offset of $-0.05 M_{\odot}$ (our result minus APOKASC) is observed, with a moderate scatter of $0.14 M_{\odot}$. This offset and dispersion are at least partly due to uncertainties in the mass-loss parameter for red giant stars.

\section{Final Sample and notes for their use}\label{subsec:Discussion}

\begin{figure*}[ht!]
\epsscale{0.8}
\plotone{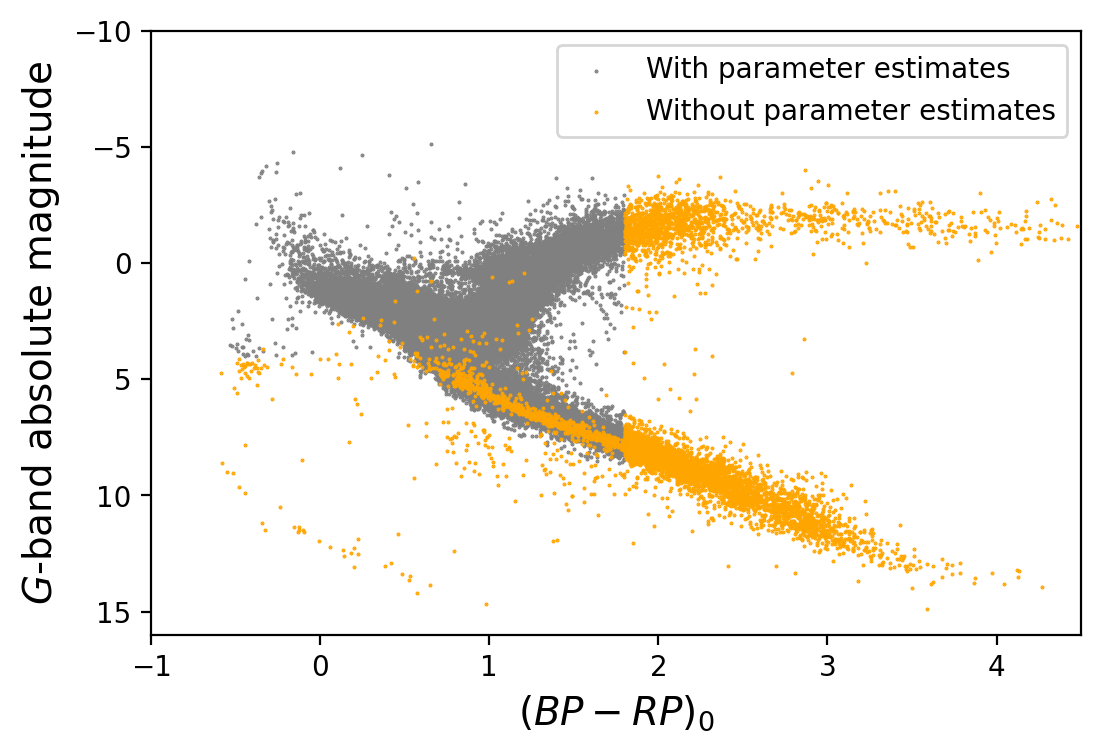}
\caption{Distribution of KIC stars (orange dots) without reliable parameter estimates on the color--$G$-band absolute magnitude diagram. The background gray dots represent stars with well-determined atmospheric and physical parameters.}
\label{fig:noproperty}
\end{figure*}

Using the methods described above, we have obtained physical parameter estimates for around 190,000 KIC stars. However, around 10,000 KIC stars still lack these estimates. An examination of the Hertzsprung-Russell (HR) diagram (see Figure \ref{fig:noproperty}) reveals that most of the stars without parameter estimates are cool dwarfs and giants with $(BP-RP)_0 \ge 1.8$, as well as hot subdwarfs and white dwarfs. These stars were excluded from the training process due to the challenges in obtaining reliable parameter estimates for them.
Additionally, a small number of main-sequence and turn-off stars lack parameter estimates because they do not have Gaia XP spectra or KIS photometry observations.
\subsection{Parameters of M-type Stars}
To obtain parameters for as many KIC stars as possible, we trained the photometric parameter relations for M-type stars using a method similar to that described in Section \ref{subsec:Parameters}. Recently, LAMOST DR10 released stellar atmospheric parameters for both M dwarf and giant stars using the LASPM pipeline \citep{2021RAA....21..202D}. We cross-matched our sample of cool stars with $(BP-RP)_0 \ge 1.8$ against the LAMOST M dwarf and giant catalog, finding over 1500 stars in (582 dwarfs and 981 
giants). Using the same training methods described in Section~\ref{subsec:Parameters}, we derived relationships between atmospheric parameters and synthesized Str{\"o}mgren photometry for both M dwarfs and giants. These relationships were then applied to over 5200 cool stars to estimate their missing atmospheric parameters. To assess the precision of our estimates, we cross-matched these stars with APOGEE DR17, finding around 100 M dwarfs and 1000 M giants in common. The comparisons indicate moderate offsets across all atmospheric parameters, with typical values around 0.10\,dex for [M/H], 120\,K for $T_{\rm eff}$, and 0.10\,dex for log~$g$. The dispersions are 0.20\,dex for both log~$g$ and [M/H], and relatively low for $T_{\rm eff}$, at about 60\,K. ue to the limited accuracy of the parameters (large offsets and dispersion), we do not proceed with isochrone-based estimates of physical parameters derived from these stellar atmospheric parameters. 
\subsection{Data Access}
In the final tables, we present data from two separate stellar catalogs: one for AFGK stars and another for M-type stars. A detailed description of the catalogs is provided in Table \ref{tab:description}. The updated KIC parameter catalogs will be publicly available at Zenodo \dataset[doi:10.5281/zenodo.14546166]{https://doi.org/10.5281/zenodo.14546166}.
\subsection{Notes for Using Data}
If one wishes to use the data from these tables, please take note of the following points:
\begin{itemize}
    \item \textbf{Parameters of M-type Stars:} Due to the relatively small size of the training set and limited data available for comparison and verification, the reliability of M-type star parameters is lower compared to that of AFGK stars. Caution is advised when using these values for further analysis.
    \item \textbf{Stellar Classification:} The classification of stellar types on the H-R diagram in this paper is based on empirical methods. Some mixtures may occur at the classification boundaries, particularly between main-sequence stars and binary stars.
    \item \textbf{Isochrone Fitting:} While the isochrone-fitting method provides reliable mass and age estimates for turn-off stars and sub-giant stars, there is greater uncertainty for other stellar types. These uncertainties should be carefully taken into account during analysis.
    \item \textbf{Surface Gravity:} In this work, stellar surface gravity was estimated using both stellar colors and isochrone fitting. Based on various checks, the accuracy of the isochrone fitting method is significantly better than that derived from stellar colors. Therefore, when surface gravity estimates are available from both methods, we recommend using the values obtained from isochrone fitting.

\end{itemize}

\section{Summary}\label{subsec:conclusion}

In this work, we have made three main improvements to the KIC catalog: (1) established a high-precision 3-D extinction map of the Kepler field, (2) obtained atmospheric parameter estimates for 97\% of KIC stars using photometric data from KIS and Gaia XP, and (3) derived stellar mass, radius, surface gravity and age estimates for these KIC stars based on their atmospheric parameters and stellar evolution models. Details of these improvements are outlined below.

\begin{itemize}
      \item First, we determined the extinction for stars in the Kepler field using the `star-pair' method and constructed a 3-D extinction map for this region. By analyzing members of four well-known open clusters within the Kepler field, we found that this new 3-D extinction map provides more accurate reddening values than those from the commonly used map by \citet{2019ApJ...887...93G}.
      
      \item By training a relationship between the photometric colors from KIS DR2, the ultra wide-band photometric colors from Gaia DR3, the photometric colors synthetically generated from the Gaia XP spectra, and the spectral stellar parameters from LAMOST DR10, we obtained atmospheric-parameter estimates for about 195,000 stars, accounting for 97$\%$ of the total number of the KIC stars. We achieved uncertainties of 0.12\,dex on [Fe/H], 100\,K on $T_{\rm eff}$, and 0.2\,dex on log $g$. 
      
      \item Using the PARSEC stellar-evolution model, we estimated the masses, radii, surface gravities, and ages of KIC stars based on their atmospheric parameters and photometric data. We then compared our mass and age estimates with values from the literature, especially stars with mass, radius, and surface gravity measurements from asteroseismology data. These comparisons indicate that our estimates achieve precisions of 0.07 $M_{\odot}$ in mass, 0.05 $R_{\odot}$ in radius, and 0.12 dex in surface gravity for dwarf stars, and 0.14 $M_{\odot}$ in mass, 0.73 $R_{\odot}$ in radius, and 0.11 dex in surface gravity for giant stars.
\end{itemize}

We summarize the methodology for each parameter estimate in Table \ref{tab:rfResult}. These results are expected to be valuable for future research on exoplanet host stars, exoplanet habitability, and asteroseismology studies. 

\begin{deluxetable*}{ccccc}[ht!]
\tablenum{3}
\tablecaption{Sample Content\label{tab:rfResult}}
\tablewidth{0pt}
\tablehead{
\colhead{Parameter Type} & \colhead{Method} & \colhead{Number of A,F,G,K stars} & \colhead{Number of M stars}& \colhead{Total}
}
\startdata
{}& KIS $U$-band photometry & 179,133 & - & 179,133 \\
{Photometric [Fe/H]}& Synthesized $v$,$b$,$y$ photometry & 189,727 & 5,252 & 194,979 \\
{}& One of the above methods & 190,226 & 5,252 & 195,478\\
Photometric $T_{\mathrm{eff}}$ & $T_{\mathrm{eff}}$-[Fe/H]-Color relation & 189,727 & 5,252 & 194,979 \\
Photometric log $g$ & log $g$ -[Fe/H]-Color relation & 179,133 & 4,285 & 183,418 \\
Age, mass, $\&$ radius & Isochrone fitting & 185,886 & \dots & 185,886
\enddata
\end{deluxetable*}

\begin{acknowledgments}

Y.H. acknowledges the supported from National Key R\&D Programme of China (Grant No. 2019YFA0405503) and the National Science Foundation of China (NSFC Grant No. 12422303). T.C.B. acknowledges partial support for this work from grant PHY 14-30152; Physics Frontier Center/JINA Center for the Evolution of the Elements (JINA-CEE), and OISE-1927130: The International Research Network for Nuclear Astrophysics (IReNA), awarded by the US National Science Foundation. K.X. acknowledge the supports of the NSFC grant No. 12403024, the Postdoctoral Fellowship Program of CPSF under Grant Number GZB20240731, the Young Data Scientist Project of the National Astronomical Data Center, and the China Post-doctoral Science Foundation No. 2023M743447.

This work presents results from the European Space Agency’s space mission Gaia (\url{https://www.cosmos.esa.int/gaia}). Gaia data are processed by the Gaia Data Processing and Analysis Consortium (\url{https://www.cosmos.esa.int/web/gaia/dpac/consortium}), which is funded by national institutions, in particular the institutions participating in the Gaia MultiLateral Agreement.

\end{acknowledgments}

\vspace{20mm}
\software{Astropy \citep{2013A&A...558A..33A,2018AJ....156..123A,2022ApJ...935..167A},  
          Dustmaps \citep{2018JOSS....3..695G}, 
          Matplotlib\citep{2007CSE.....9...90H},
          multiprocessing\citep{mckerns2012building}
          NumPy\citep{2020Natur.585..357H}, 
          pandas\citep{2022zndo...3509134T},
          Scikit-learn\citep{2011JMLR...12.2825P},
          SciPy\citep{2020NatMe..17..261V}
          Topcat\citep{2005ASPC..347...29T}}

\bibliography{paperAfterRevise}{}
\bibliographystyle{aasjournal}

\appendix

\section{Extinction Map Comparison}

Here we summarize the adopted parameters for the four open clusters in the Kepler field (Table~\ref{tab:openCluster}), compare the reddening values obtained based on our 3-D extinction map of the Kepler field with the values of reddening adopted by \citet[][see Figure~\ref{fig:compareEBV}]{2019ApJ...887...93G}, and compare the reddening values from our 3-D extinction map for member stars in the four open clusters in the Kepler field with those adopted by \citet[][see Figure~\ref{fig:interpolation}]{2019ApJ...887...93G}.

\begin{deluxetable}{cccccccc}[h!]
\tablenum{A1}
\tablecaption{Parameters of open clusters in the Kepler field\label{tab:openCluster}}
\tablewidth{0pt}
\tablehead{
\colhead{NGC} & \colhead{Distance$^a$} & \colhead{RA$^a$} &
\colhead{DEC$^a$} & \colhead{$\text{PM}_{RA}^a$} & \colhead{$\text{PM}_{DEC}^a$} & \colhead{$E(B-V)$} & \colhead{Age}\\
\colhead{Number} & \colhead{(kpc)} & \colhead{(°)} &
\colhead{(°)} & \colhead{(mas/yr)} & \colhead{(mas/yr)} & \colhead{(mag)} & \colhead{(Gyr)}
}
\decimals
\startdata
NGC 6791 & 4.2 & 290.22 & 37.77 & $-$0.42 & $-$2.28 &0.10$^a$/0.14$^b$ & 8.30$^c$/11.7$^b$\\ 
NGC 6811 & 1.1 & 294.34 & 46.36 & $-$3.35 & $-$8.80 & 0.074$^d$/0.072$^b$ &1.05$^e$/1.21$^b$\\
NGC 6819 & 2.5 & 295.32 & 40.19 & $-$2.90 & $-$3.87 & 0.16$^f$/0.14$^b$ & 2.38$^g$/2.16$^b$\\
NGC 6866 & 1.4 & 300.98 & 44.16 & $-$1.37 & $-$5.76 & 0.13$^h$/0.13$^b$ & 0.78$^h$/0.90$^b$\\
\enddata
\tablecomments{References: a-\citet{2023AA...673A.114H}, b-this work, c-\citet{2021AA...649A.178B}, d-\citet{2013AJ....145....7J}, e-\citet{2016ApJ...831...11S}, f-\citet{2014AJ....148...51A}, g-\citet{2016AJ....151...66B}, h-\citet{2019AA...623A.108B}}
\end{deluxetable}

\renewcommand\thefigure{\Alph{section}\arabic{figure}}    

\setcounter{figure}{0}  

\begin{figure*}[h!]
\gridline{\fig{/starpair/compare_g19.png}{0.7\textwidth}{}
}

\caption{Comparison of reddening values between those derived using the 'star-pair' method in this work and those from the 3-D extinction map of \citet{2019ApJ...887...93G}. $\Delta E (B-V)$ represents the difference between the reddening values obtained from the `star-pair' method and those from the 3-D extinction map of \citet{2019ApJ...887...93G}. The red-dashed line represents the zero level. The mean and scatter of this difference are indicated in the bottom-right corner.
\label{fig:compareEBV}}
\end{figure*}

\begin{figure*}[h!]
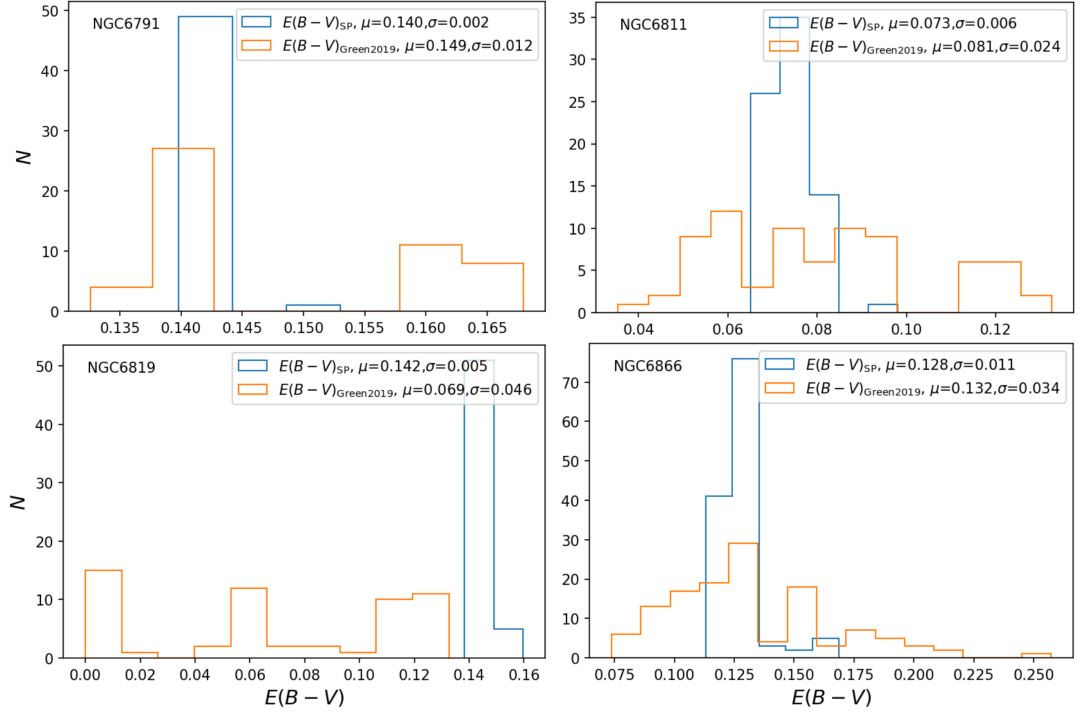

\gridline{\fig{/starpair/ebv_cluster}{0.8\textwidth}{}
          }
\caption{Reddening values for member stars of four open clusters (NGC 6791, NGC 6811, NGC 6819, and NGC 6866) in
the Kepler field. The blue lines represent $E (B -V)$ given by our 3-D extinction map, while the orange lines indicate the $E (B - V)$ from the map of \citet{2019ApJ...887...93G}. The mean and dispersion of these distributions are marked in upper right of each panel. \label{fig:interpolation}}
\end{figure*}

\section{Supplementary Metallicity Comparison Test}

Here we provide comparisons of the photometric metallicity estimates from the KIS photometry for turn-off stars and binary stars with those from APOGEE DR17 (Figure~\ref{fig:turnoffkis}), and photometric metallicity estimates from
Str{\"o}mgren photometry synthesized from the Gaia XP spectra for turn-off stars and binary stars with those from APOGEE DR17 (Figure~\ref{fig:turnoffgaia}).

\setcounter{figure}{0}  
\begin{figure}[ht!]
\centering  
\includegraphics[width=0.8\textwidth]{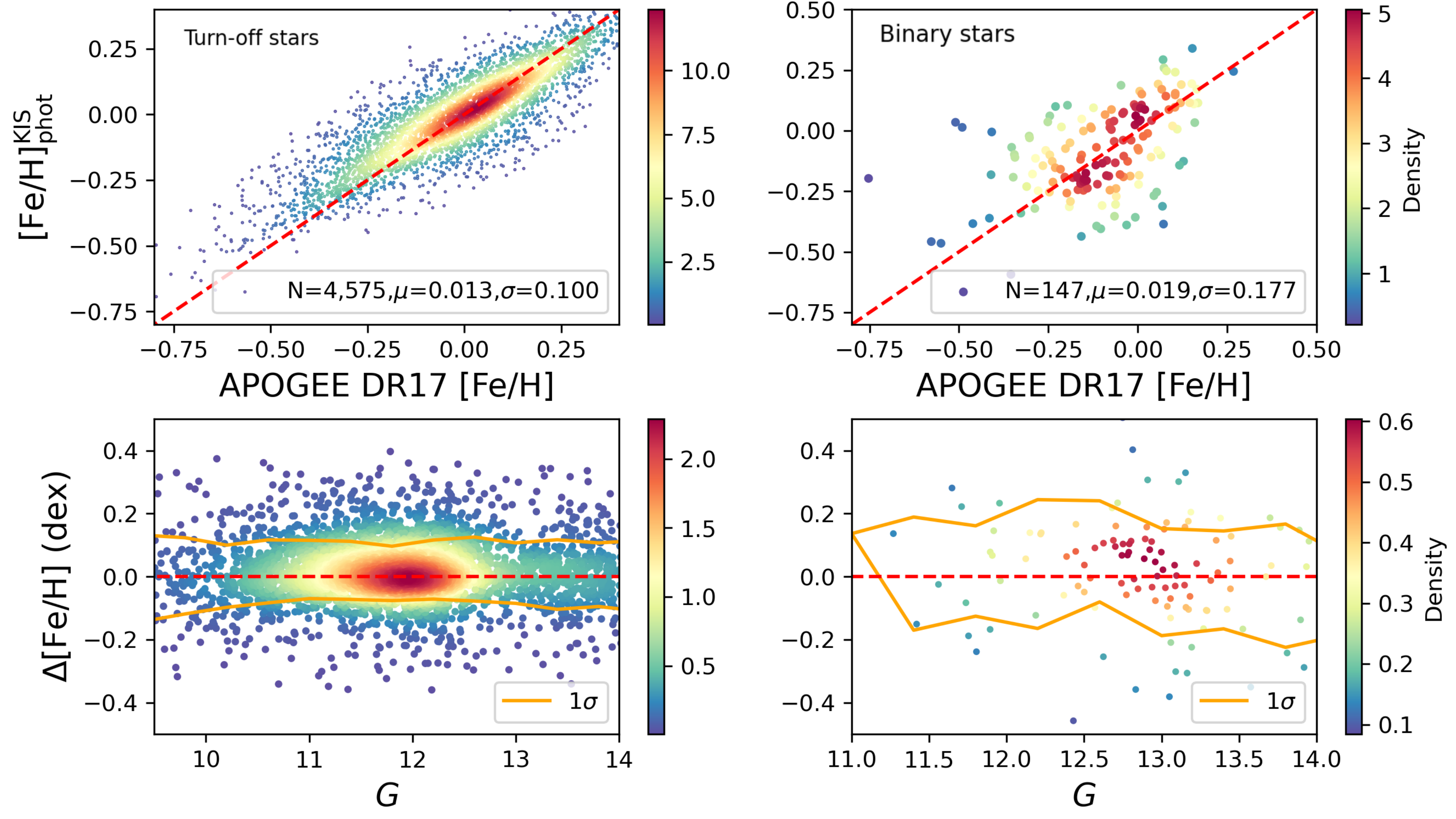}
\caption{Comparisons of photometric metallicity estimates from the KIS photometry with those from the APOGEE DR17 for the turn-off stars (left column of panels) and binary stars (right column of panels).
The red-dashed lines in the upper panels are the one-to-one lines. The numbers of stars, mean offset, and scatter are provided at the lower right of each panel. The lower panels are the residuals of metallicity differences (${\rm [Fe/H]_{~phot}}-{\rm [Fe/H]_{APOGEE}}$) as a function of Gaia $G$ magnitude. The red-dashed lines in the lower panels represent the zero level.  The golden lines in the lower panels represents the $1\sigma$ scatter. The color bar to the right of each panel codes the number density of stars in the panel. 
\label{fig:turnoffkis}}
\end{figure}

\begin{figure}[ht!]
\centering  
\includegraphics[width=0.8\textwidth]{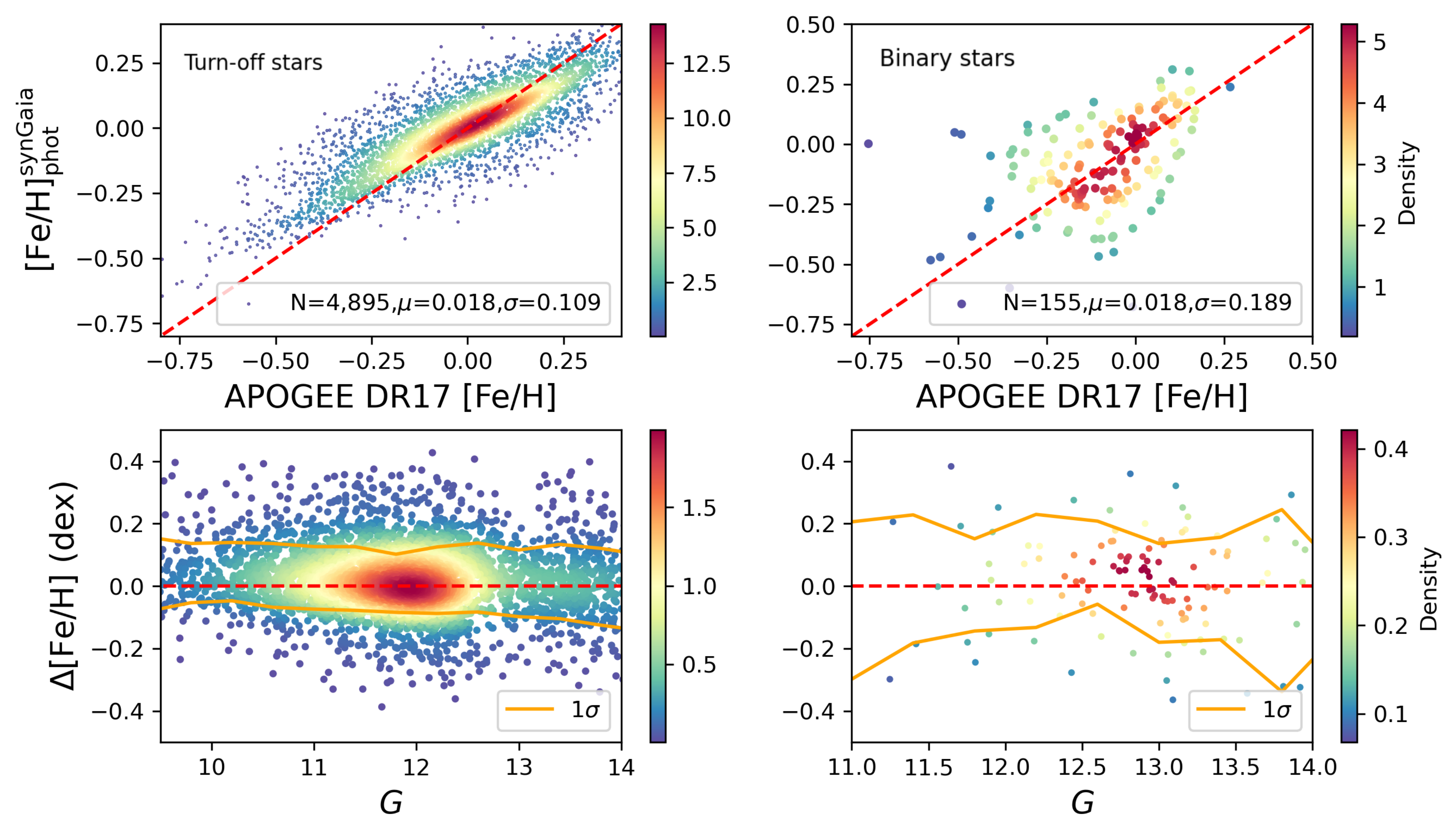}
\caption{Similar to Figure~\ref{fig:turnoffkis}, but for photometric metallicity estimates from the Str{\"o}mgren photometry synthesized from the Gaia XP spectra. 
\label{fig:turnoffgaia}}
\end{figure}

\section{Catalog Description}

Here we provide a detailed listing of the contents of our final catalog, including the input quantities
and their sources, as well as the derived quantities and estimates of stellar atmospheric parameters from LAMOST DR10 and APOGEE DR17.  The updated KIC parameter catalogs will be publicly available at \dataset[doi:10.5281/zenodo.14546166]{https://doi.org/10.5281/zenodo.14546166}

\begin{deluxetable}{cc}[ht!]
\tablenum{C1}
\tablecaption{Catalog Description\label{tab:description}}
\tablewidth{0pt}
\tablehead{
\colhead{Column name} & \colhead{Description}
}
\startdata
kepid & ID of the KIC star\\
SourceID & Source ID of Gaia DR3\\
degree$\_$ra & RA of KIC star\\
degree$\_$dec & DEC of KIC star\\
$G$ & $G$-band photometry from Gaia DR3\\
$G$$\_$err & Uncertainty of $G$-band photometry from Gaia DR3\\
$BP$ & $BP$-band photometry from Gaia DR3\\
$BP$$\_$err & Uncertainty of $BP$-band photometry from Gaia DR3\\
$RP$ & $RP$-band photometry from Gaia DR3\\
$RP$$\_$err & Uncertainty of $RP$-band photometry from Gaia DR3\\
$BP-RP$ & $BP-RP$ color from Gaia DR3\\
$v$ & $v$-band photometry synthesized from Gaia DR3 XP spectra\\
$v$$\_$err & Uncertainty of $v$-band photometry synthesized from Gaia DR3 XP spectra\\
$b$ & $b$-band photometry synthesized from Gaia DR3 XP spectra\\
$b$$\_$err & Uncertainty of $b$-band photometry synthesized from Gaia DR3 XP spectra\\
$y$ & $y$-band photometry synthesized from Gaia DR3 XP spectra\\
$y$$\_$err & Uncertainty of $y$-band photometry synthesized from Gaia DR3 XP spectra\\
$U$ & $U$-band photometry from KIS DR2\\
$U$$\_$err & Uncertainty of $U$-band photometry from KIS DR2\\
$E(BP-RP)$ & $BP-RP$ excess from our 3-D extinction map\\
FeH$\_$KIS$\_$PHOT/MH$\_$KIS$\_$PHOT$^a$& [Fe/H]/[M/H] from KIS colors\\
FeH$\_$KIS$\_$PHOT$\_$err/MH$\_$KIS$\_$PHOT$\_$err$^a$ & Uncertainty of [Fe/H]/[M/H] from KIS colors\\
FeH$\_$GaiaSyn$\_$PHOT/MH$\_$GaiaSyn$\_$PHOT$^a$ & [Fe/H]/[M/H] from Gaia synthesized colors\\
FeH$\_$GaiaSyn$\_$PHOT$\_$err/MH$\_$GaiaSyn$\_$PHOT$\_$err$^a$ & Uncertainty of [Fe/H]/[M/H] from Gaia synthesized colors\\ 
$T_{\mathrm{eff}}$$\_$PHOT & Photometric Effective temperature\\
$T_{\mathrm{eff}}$$\_$PHOT$\_$err & Photometric Uncertainty of effective temperature\\
log $g$$\_$PHOT & Photometric Surface gravity\\
log $g$$\_$PHOT$\_$err & Photometric Uncertainty of surface gravity\\
Age$\_$ISO$^b$ & Age by isochrone-fitting method\\
Age$\_$ISO$\_$low$^b$ & 16th percentile of age posterior by isochrone-fitting method\\
Age$\_$ISO$\_$up$^b$ & 84th percentile of age posterior by isochrone-fitting method\\
$M$$\_$ISO & Mass$^b$ by isochrone-fitting method\\
$M$$\_$ISO$\_$err$^b$ & Uncertainty of mass fitted by isochrone-fitting method\\
log $g$$\_$ISO$^b$ & Surface gravity by isochrone-fitting method\\
log $g$$\_$ISO$\_$err$^b$ & Uncertainty of surface gravity by isochrone-fitting method\\
$T_{\mathrm{eff}}$$\_$ISO$^b$ & Effective temperature by isochrone-fitting method\\
$T_{\mathrm{eff}}$$\_$ISO$\_$err$^b$ & Uncertainty of effective temperature by isochrone-fitting method\\
log $L$$\_$ISO$^b$ & Luminosity by isochrone-fitting method\\
log $L$$\_$ISO$\_$err$^b$ & Uncertainty of luminosity by isochrone-fitting method\\
$R$$\_$ISO$^b$ & Radius by isochrone-fitting method\\
$R$$\_$ISO$\_$err$^b$ & Uncertainty of radius fitted by isochrone-fitting method\\
\enddata
\tablecomments{a:[Fe/H] for A, F, G and K type stars and [M/H] for M type dwarfs.\\
b: Only for A, F, G and K type stars.}
\end{deluxetable}

\begin{deluxetable}{cc}[ht!]
\startlongtable
\clearpage
\tablenum{B1}
\tablewidth{0pt}
\tablehead{
\colhead{Column name} & \colhead{Description}
}
\startdata
FeH$\_$LAMOST/MH$\_$LAMOST$^{a}$ & [Fe/H]/[M/H] from LAMOST DR10\\
FeH$\_$LAMOST$\_$err/MH$\_$LAMOST$\_$err$^{a}$ & Uncertainty of [Fe/H]/[M/H] from LAMOST DR10\\
$T_{\mathrm{eff}}$$\_$LAMOST & Effective temperature from LAMOST DR10\\
$T_{\mathrm{eff}}$$\_$LAMOST$\_$err & Uncertainty of effective temperature from LAMOST DR10\\
log $g$$\_$LAMOST & Surface gravity from LAMOST DR10\\
log $g$$\_$LAMOST$\_$err & Uncertainty of surface gravity from LAMOST DR10\\
FeH$\_$APOGEE & [Fe/H] from APOGEE DR17\\
FeH$\_$APOGEE$\_$err & Uncertainty of [Fe/H] from APOGEE DR17\\
$T_{\mathrm{eff}}$$\_$APOGEE & Effective temperature from APOGEE DR17\\
$T_{\mathrm{eff}}$$\_$APOGEE$\_$err & Uncertainty of effective temperature from APOGEE DR17\\
log $g$$\_$APOGEE & Surface gravity from APOGEE DR17\\
log $g$$\_$APOGEE$\_$err & Uncertainty of surface gravity from APOGEE DR17\\
MH$\_$APOGEE & [M/H] from APOGEE DR17\\
MH$\_$APOGEE$\_$err & Uncertainty of [M/H] from APOGEE DR17\\
pmra & Proper motion in R.A. direction from Gaia DR3\\
pmra$\_$err & Uncertainty of proper motion in R.A. direction from Gaia DR3\\
pmdec & Proper motion in decl. direction from Gaia DR3\\
pmdec$\_$err & Uncertainty of proper motion in decl. direction from Gaia DR3\\
rgeo & Geometric distance from \cite{2021AJ....161..147B}\\
rgeo$\_$low & 16th percentile of the geometric distance posterior from \cite{2021AJ....161..147B}\\
rgeo$\_$up & 84th percentile of the geometric distance posterior from \cite{2021AJ....161..147B}\\
rpgeo & Photogeometric distance from \cite{2021AJ....161..147B}\\
rpgeo$\_$low & 16th percentile of the photogeometric distance posterior from \cite{2021AJ....161..147B}\\
rpgeo$\_$up & 84th percentile of the photogeometric distance posterior from \cite{2021AJ....161..147B}\\
StarType & Type of KIC star, including TO (turn-off), Giant, MS (main sequence), Binary, HS (hot star),\\
{} & gM (M-type giant) and dM (M-type dwarf) \\
\enddata
\tablecomments{a: [Fe/H] for A, F, G and K type stars and [M/H] for M type stars.}
\end{deluxetable}


\end{document}